\documentclass[preprint2]{aastex}

\newcommand{\h}{^{\rm h}}
\newcommand{\m}{^{\rm m}}
\newcommand{\ratio}{_{3-2/1-0}}

\usepackage[usenames]{color}

\shorttitle{GMC Evolutions in M33}

\shortauthors{Miura et al.}

\begin{document}


\title{GMC Evolutions in the Nearby Spiral Galaxy M33}


\author{Rie E. Miura\altaffilmark{1,2,3}, Kotaro Kohno\altaffilmark{3,4}, Tomoka Tosaki\altaffilmark{5}, Daniel Espada\altaffilmark{1}, Narae Hwang \altaffilmark{1}, Nario Kuno\altaffilmark{6}, Sachiko K. Okumura\altaffilmark{1,2}, Akihiko Hirota\altaffilmark{6}, Kazuyuki Muraoka\altaffilmark{8}, Sachiko Onodera\altaffilmark{6,9}, Tetsuhiro Minamidani\altaffilmark{10}, Shinya Komugi\altaffilmark{1,11}, Kouichiro Nakanishi\altaffilmark{1,11,12}, Tsuyoshi Sawada\altaffilmark{1,11}, Hiroyuki Kaneko\altaffilmark{6,12,13} \and Ryohei Kawabe\altaffilmark{6,11}
}
\altaffiltext{1}{National Astronomical Observatory of Japan, 2-21-1 Osawa, Mitaka, Tokyo, 181-8588, Japan}
\email{rie.miura@nao.ac.jp}
\altaffiltext{2}{Department of Astronomy, The University of Tokyo, Hongo, Bunkyo-ku, Tokyo, 133-0033, Japan}
\altaffiltext{3}{Institute of Astronomy, School of Science, The University of Tokyo, Osawa, Mitaka, Tokyo 181-0015, Japan}
\altaffiltext{4}{Research Center for Early Universe, School of Science, The University of Tokyo, Hongo, Bunkyo, Tokyo, 113-0033, Japan}
\altaffiltext{5}{Joetsu University of Education, Yamayashiki-machi, Joetsu, Niigata, 943-8512, Japan} 
\altaffiltext{6}{Nobeyama Radio Observatory, Minamimaki, Minamisaku, Nagano,384-1805, Japan}
\altaffiltext{7}{Department of Mathematical and Physical Sciences, Faculty of Science, Japan Woman's University, Mejirodai 2-8-1, Bunkyo, Tokyo, 112-8681, Japan }
\altaffiltext{8}{Osaka Prefecture University, Gakuen 1-1, Sakai, Osaka, 599-8531, Japan}
\altaffiltext{9}{Department of Physics, Meisei University, Hino, Tokyo, 191-8506, Japan}
\altaffiltext{10}{Department of Physics, Faculty of Science, Hokkaido University, N10W8, Kita-ku, Sapporo 060-0810, Japan}
\altaffiltext{11}{Joint ALMA Observatory, Alonso de Cordova 3107, Vitacura 763-0355, Santiago de Chile}
\altaffiltext{12}{The Graduate University for Advanced Studies (Sokendai), 2-21-1 Osawa, Mitaka, Tokyo 181-0015, Japan}
\altaffiltext{13}{Institute of Physics, University of Tsukuba, 1-1-1 Tennodai, Tsukuba, Ibaraki, 305-8571, Japan}








\begin{abstract}
We present a Giant Molecular Cloud (GMC) catalog toward M33, containing 71 GMCs in total, based on wide field and high sensitivity CO($J=3-2$) observations with a spatial resolution of 100\,pc using the ASTE 10\,m telescope. Employing archival optical data, we identify 75 young stellar groups (YSGs) from the excess of the surface stellar density, and estimate their ages by comparing with stellar evolution models. A spatial comparison among the GMCs, YSGs, and H{\sc ii} regions enable us to classify GMCs into four categories: Type~A showing no sign of massive star formation (SF), Type~B being associated only with H{\sc ii} regions, Type~C with both H{\sc ii} regions and $<10$\,Myr-old YSGs and Type-D with both H{\sc ii} regions and 10--30 Myr YSGs. Out of 65 GMCs (discarding those at the edges of the observed fields), 1 (1\%), 13 (20\%), 29 (45\%), and 22 (34\%) are Types~A, B, C, and D, respectively. We interpret these categories as stages in
a GMC evolutionary sequence. Assuming that the timescale for each evolutionary stage is proportional to the number of GMCs, the lifetime of a GMC with a mass $>10^5\,M_{\odot}$ is estimated to be 20--40\,Myr. In addition, we find that the dense gas fraction as traced by the CO($J=3-2$)/CO($J=1-0$) ratio is enhanced around SF regions. This confirms a scenario where dense gas is preferentially formed around previously generated stars, and will be the fuel for the next stellar generation. In this way, massive SF gradually propagates in a GMC until gas is exhausted.

\end{abstract}

\keywords{Galaxies: individual (M33) --- ISM: clouds --- ISM:molecules --- Submillimeter: ISM}

\section{Introduction}
\label{p3:intro}
Observations in our Galaxy and in the Local Group of Galaxies (LGGs) have revealed that Giant Molecular Clouds (GMCs) have masses of about $10^4$ to 10$^7$\,M$_{\odot}$ and sizes of  50 to several hundred of parsecs \citep[e.g.,][]{1993prpl.conf..125B,2010ARA&A..48..547F}, which are often found associated with H{\sc ii} regions and OB associations \citep{2003ARA&A..41...15M}, tracers of massive star formation (SF).
Massive stars, which are commonly found clustered in OB associations \citep{2003ARA&A..41...57L}, are short-lived (less than a few 10\,Myr),
and thus spend most of their lifetime within the cluster in which they were formed. 
This suggests that GMCs are the principal sites of massive SF.
However, how SF occurs in a GMC, the so-called ``GMC evolution'', has remained poorly understood because it has been difficult to perform molecular gas observations with a combination of high resolution, sensitivity and wide field to enable studies of a sufficiently large sample of GMCs.
Such studies in LGGs have the potential to reveal the density distribution and kinematics at GMC scales, as well as its relation to ongoing SF.

Several studies focusing on GMC evolution in the Large Magellanic Clouds (LMC) have been performed \citep[e.g.,][]{1999PASJ...51..745F,2001PASJ...53..971M,2001PASJ...53..985Y,2009ApJS..184....1K}.
The GMCs in the LMC are classified into three types according to their associated massive SF activities: Type~{\sc i} showing no signature of massive SF, Type~{\sc ii} being associated with relatively small H{\sc ii} regions, and Type~{\sc iii} with
both H{\sc ii} regions and young (less than 10\,Myr old) stellar clusters \citep{2001PASJ...53..985Y,2009ApJS..184....1K}. 
While in general the positions of young clusters are correlated with GMCs, the clusters older than 10\,Myr have a weaker or no correlation with the GMCs \citep{2009ApJS..184....1K}. 
This classification was interpreted to be a template for the GMC evolutionary stages, and the typical lifetime of GMCs with masses larger than $5 \times 10^4$\,M$_{\odot}$ is estimated to be 20--30\,Myr  \citep{2007prpl.conf...81B,2009ApJS..184....1K}.
Analyses of the kinetic temperature and density of several GMCs in LMC shows that  they increase generally as the GMCs evolve \citep{2008ApJS..175..485M,2008A&A...482..197P, 2010PASJ...62...51M, 2011AJ....141...73M}. 
This confirms that the denser and warmer molecular gas is directly linked to SF activities.
Because these studies are only limited to several specific GMCs in one irregular galaxy, it is still necessary to investigate the properties of GMCs in the different environments present in other galaxies, in order to complete our current knowledge of templates for GMC evolution.

M33 is one of the best laboratories to study GMC evolutions under different environments due to its proximity \citep[distance of 840\,kpc;][]{2001ApJ...553...47F}, favorable inclination \citep[$i = 51^{\circ}$;][]{2000MNRAS.311..441C}, as well as the existence of spiral structure and massive SF regions. M33 is the second closest spiral galaxy after M31, but M33 is less inclined, which is ideal to resolve gas components or individual stars with little contamination along the line of sight.
Also, studying GMCs over a disk galaxy enable us to compare all GMCs at the same distance, unlike in our Galaxy.

Surveys of $^{12}$CO($J=1-0$) and $^{12}$CO($J=2-1$) emission in M33 to investigate the relation between molecular gas and massive SF have been conducted by several authors \citep{2003ApJS..149..343E, 2004ApJ...602..723H, 2007ApJ...661..830R, 2010A&A...522A...3G,2011PASJ...63.1171T, 2012A&A...542A.108G}.
\citet{2003ApJS..149..343E} have found that more than two-thirds of the GMCs in M33 are associated with H{\sc ii} regions within 50\,pc, which is similar to the proportion in LMC \citep{2001PASJ...53..985Y,2009ApJS..184....1K}.
On the other hand, the correlation between the molecular gas surface densities traced by CO($J=1-0$) emission and that of SF rate (SFR), the so-called Kennicutt--Schmidt Law \citep[hereafter K-S law;][]{1959ApJ...129..243S,1989ApJ...344..685K} in a pixel-to-pixel analysis is found to be weaker at 80\,pc resolutions \citep{2010ApJ...722L.127O} than at lower resolutions \citep[200\,pc;][]{2004ApJ...602..723H}.
This suggests that the GMCs traced by CO($J=1-0$) are well correlated with the massive SF sites, but their peaks are offset from each other. 

Recently, the K-S law using CO($J=3-2$) in M33 has shown a tight correlation at 100\,pc resolution, unlike CO($J=1-0$) (Onodera et al. 2013, PASJ, accepted). In fact, with higher resolution data, \citet{2007ApJ...664L..27T} have found that the
CO($J=3-2$) emission is spatially better correlated to massive SF sites than
CO($J=1-0$) in the most luminous giant H{\sc ii} region (GHR) in M33, NGC~604.  This supports that the dense and warm gas traced by CO($J=3-2$) is more closely linked to the SF sites, as previously argued in four Virgo galaxies in \citet{2009ApJ...693.1736W}.
These results are interpreted as higher kinetic temperatures and densities in GMCs found close to GHRs,  in agreement with large velocity gradient (LVG) analysis with multi-$J$ CO transition data by \citet{1997ApJ...483..210W} and \citet{2012PASJ...64....3M}.

As part of the Nobeyama Radio Observatory (NRO) M33 All-disk survey of Giant Molecular Clouds (MAGiC) project \citep{2011PASJ...63.1171T}, 
here we present a catalog of M33 GMCs using new CO($J=3-2$) maps with $25\arcsec$ (corresponding to 100\,pc) resolution. 
In order to investigate the relation of the GMCs with massive SF, we also present a ``young stellar group (YSG)'' catalog.
In this work, YSGs can be clusters (typical sizes of 15\,pc), OB associations (typical sizes of $\sim$15--100\,pc),  and star complexes (sizes of a few hundred pc) \citep[e.g.,][]{1996A&A...314...51B,2005PASRB...4...75I}.
There are a number of catalogs for clusters, OB associations, and star complexes in M33 published so far (e.g., \citealp{2007AJ....134..447S}, hereafter SM; \citealp{2007AJ....134.2168P}; \citealp{2010ApJ...720.1674S}; \citealp{1980ApJS...44..319H}; \citealp{2005PASRB...4...75I}).
However, the number of star clusters whose ages have been estimated is limited, and the methods to derive the ages are not homogenous (SM and references therein; \citealp{2009ApJ...700..103P}, hereafter PPL; \citealp{2009ApJ...699..839S}, hereafter SSGH).
We also aim to estimate their ages in a homogeneous manner using the stellar photometry catalog provided by \citet{2006AJ....131.2478M}. 

The outline of this paper is as follows.
The observations and data reduction are summarized in Section~\ref{p3:obs}.
In Section~\ref{p3:res} we explain the general procedure for identifying GMCs, as well as the identification of YSGs and the estimation of their ages. 
We classify GMCs as a function of the age of YSGs associated with them. 
We also quantify the SF activities and dense gas fractions in the GMCs using the extinction-corrected H$\alpha$ data and the CO($J=3-2$)/CO($J=1-0$) line ratio, $R\ratio$, and then show the variety of physical states of the identified GMCs. 
In Section~\ref{p3:disc}, we discuss the relationship between the properties of these clouds
and their evolutionary stages. Finally, we interpret our results in
the context of a continuous SF in GMCs.
The summary of this paper is provided in Section~\ref{p3:sum}.

\section{Observations and Data Reduction}
\label{p3:obs}
We describe the data that is used throughout this paper, including the CO($J=3-2$) data to identify the GMCs, the CO($J=1-0$) data
to derive $R\ratio$, and  the optical and infrared data to identify the YSG and H{\sc ii} regions in M33.

\subsection{ASTE CO($J=3-2$) Data}
\label{p3:sec321}
We chose eight regions covering  the northern and southern spiral arms as well as the galaxy center, where the CO($J=1-0$) emission is prominent \citep{2011PASJ...63.1171T}.
The CO($J=3-2$) (rest frequency: 345.796\,GHz) mapping observations were performed between June and November 2011, using the Atacama Submillimeter Telescope Experiment \citep[ASTE;][]{2004SPIE.5489..763E,2008SPIE.7012E...6E,2005ASPC..344..242K} 10-m dish equipped with the 345-GHz side band separating (2SB) SIS receiver CATS345 \citep[CArtridge-Type Sideband-separating receiver for ASTE 345-GHz band;][]{2008stt..conf..281I}.
An XF-Type~Digital spectrometer MAC \citep{2000SPIE.4015...86S} was used to cover a velocity width of 445 km\,s$^{-1}$ with a velocity resolution of 2.5\,km\,s$^{-1}$ at 345 GHz.
The On-The-Fly (OTF) mapping technique \citep[][]{2007A&A...474..679M}, based on scanning smoothly and rapidly along one direction across a rectangular map, was employed to obtain the CO($J=3-2$) data.
The observed area covers in total about 80\,\% of the CO($J=1-0$) based molecular gas mass in \citet{2011PASJ...63.1171T}.

The observations and the data calibration were performed for each region as follows.
The half power beam width (HPBW) of the ASTE 10-m telescope is 22$\arcsec$ at 345 GHz.
The chopper-wheel technique was employed to calibrate the antenna temperature $T_A^{\ast}$ and 
the final data has been converted into units of main beam brightness temperature ($T_{\rm mb} \equiv T_A^{\ast}/\eta_{\rm mb}$), where $\eta_{\rm mb}$ is the main beam efficiency, measured  to be $0.6\,\pm\,0.1$ using the standard source Orion KL at ($\alpha_{\rm J2000}$, $\delta_{\rm J2000}$)=(05$\h$35$\m$14\fs5, 05\arcdeg22\arcmin29\farcs6). 
Hereafter, all the CO intensity measurements are specified in $T_{\rm mb}$.
The typical system temperatures in a single side band were usually less than 200\,K. 
The absolute pointing accuracy was verified by observing the CO($J=3-2$) emission of o-Cet at 1--1.5 hr intervals, and it was kept better than 4$\arcsec$ rms. 
The absolute intensity stability was also monitored using o-Cet during the different observing runs and was found to be stable within $\sim16$\,\%.

The data reduction was carried out with the Nobeyama OTF Software Tools for Analysis and Reduction package \citep[NOSTAR,][]{2008PASJ...60..445S}. 
The ``scanning
noise'' was removed by combining scans using the PLAIT
algorithm as described by \citet{1988A&A...190..353E}.
The data have been convolved with a Bessel-Gauss function in the spectral domain to create the data cube. The final grid spacing is 8\farcs0 and the angular resolution is 25\arcsec.
The final data cubes for the eight regions are characterized by an rms noise of $\sigma_{\rm ch}$ = 16--32\,mK in a velocity resolution of 2.5\,km\,s$^{-1}$, achieved after 87 hr of total integration. We briefly summarize in Table \ref{p3:tab1} the observational and data reduction parameters for the eight observed regions, including their central positions, sizes, integration times, and rms for a 2.5\,km\,s$^{-1}$ channel.

\begin{figure}
\epsscale{1.0}
\plotone{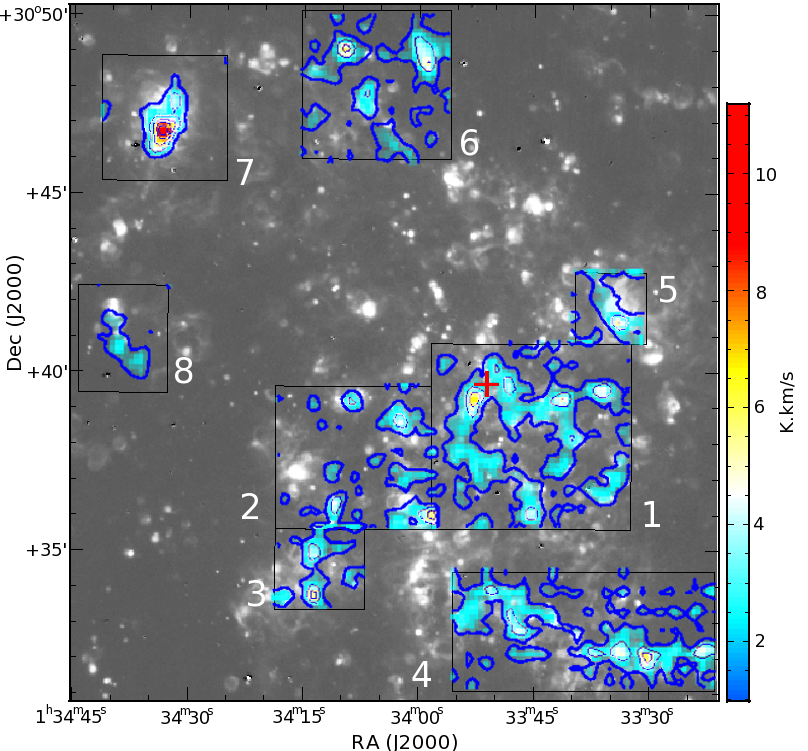}
\caption{The CO($J=3-2$) integrated intensity map with 25$\arcsec$ resolution ($\sim$ 100\,pc), overlaid on the H$\alpha$ image.
Color range spans from 1 to 11\,${\rm K\,km\,s}^{-1}$ and contour levels are 1, 3, 5, 7, and 9\,${\rm K\,km\,s}^{-1}$. 
The rectangular boxes indicate the eight observed regions in CO$(J=3-2)$, and the number identifies each region in Table \ref{p3:tab1}. The cross symbol in Box 1 represents the galaxy center.
\label{p3:fig2}}
\end{figure}

Figure~\ref{p3:fig2} shows the obtained integrated intensity map over the velocity range $V_{\rm LSR}=-274$\,km\,s$^{-1}$ to $-94$\,km\,s$^{-1}$, where emission is $>$ 2\,$\sigma_{\rm ch}$, overlaid on the H$\alpha$ image (\S~\ref{p3:2.2}). The rms noise level of the integrated intensity map ($\sigma_{\rm mom}$) is 0.38\,K\,km\,s$^{-1}$ on average.
The noise level of the CO($J=3-2$) integrated intensity map is calculated using $\sigma_{\rm mon}\equiv\sigma_{\rm ch}\sqrt{N}\delta v$, where $N$ is the number of integrated channels and $\delta v$ is the velocity resolution of a channel (2.5\,km\,s$^{-1}$).
The observed fields cover the molecular disk up to the galactic radius of $\sim$5\,kpc.
Note that we also present CO integrated intensity maps for each individual GMC in Section \ref{sec311} and the Appendix.

\subsection{NRO CO($J=1-0$) Data}
\label{p3:sec322}
The CO($J=1-0$) emission was observed in a $31\arcmin \times 36\arcmin$ area (corresponding to 7.6 kpc $\times$ 8.8 kpc) toward the disk of M33 using the NRO 45-m telescope \citep{2011PASJ...63.1171T}.
The data reduction was also carried out with the NOSTAR reduction package.
The angular resolution of the final map was 19\farcs3 and the grid spacing 7\farcs5.
Further processing was done with MIRIAD \citep{1995ASPC...77..433S}, in a manner similar to the CO($J=3-2$) map.
We have convolved the CO($J=1-0$) map to a
common angular resolution of 25$\arcsec$ and re-gridded it to 8\farcs0 per pixel.
The convolved map had an rms noise of 87\,mK for the velocity resolution of 2.5\,km\,s$^{-1}$.
We have also created CO($J=1-0$) integrated intensity maps for each GMC (see Appendix), integrating over the same velocity range as used for the individual CO($J=3-2$) integrated intensity maps.

\subsection{Optical and Infrared Data}
\label{p3:2.2}
We have used the photometric optical dataset from the  {\it UBVI} ground-based survey
of local star-forming galaxies with the Kitt Peak National Observatory 4\,m telescope presented in \citet{2006AJ....131.2478M},
and available on their ftp site\footnote{ftp://ftp.lowell.edu/pub/massey/lgsurvey/}. It contains a total of 146,622 stars in a field of 0.8 deg$^2$ centered on M33, which fully covers the regions that were observed in CO($J=3-2$) and CO($J=1-0$).
The photometry in {\it BVI} bands is less than 1\,\% above 21.1 mag and even less than 10\,\% at about 23~mag. Refer to \citet{2006AJ....131.2478M}
for further details about the observation, data reduction and photometric analysis.

We have also compiled available H{\sc ii} region catalogs \citep{1987A&A...174...28C, 1997PASP..109..927W,1999PASP..111..685H}. The H$\alpha$ image from \citet{2000ApJ...541..597H} is used to check the accuracy of the positions and extents of the M33 individual H{\sc ii} regions in these catalogs.
Refer to \citet{2000ApJ...541..597H} for more details on the observations and reduction process.

Finally, we have used the 24\,$\micron$ {\it Spitzer} data presented in \citet{2010ApJ...722L.127O} to correct the extinction in the H$\alpha$ emission. 
The final combined mosaic image is approximately $1.1\times1.2$\,deg$^{2}$, and both the point-spread-function FWHM and grid sizes are set to $5\farcs7$.
For more details on the reduction process, refer to \citet[][and references therein]{2010ApJ...722L.127O}.
Note that \citet{2010ApJ...722L.127O} have performed a pixel-by-pixel analysis and have not applied a local background subtraction to
create the extinction-corrected data by combining
the H$\alpha$ image and the $24\,\micron$ data. The lack of background subtraction may cause
a systematic shift in the extinction-corrected H$\alpha$ luminosity \citep[e.g.][]{2009ApJ...703.1672K}.
In this
paper, instead, we present a new approach in which the H$\alpha$ and $24\,\micron$ luminosities are measured with circular apertures of several 100 pc radius and with local background subtraction applied (see Section\,\ref{p3:association}).
We use the catalog of $24\,\micron$ sources in \citet{2007A&A...476.1161V}, which is the only reference listing 24\,$\micron$ sources to date, for the sake of comparing the positions of H{\sc ii} regions. 
\section{Results}
\label{p3:res}
\subsection{The CO($J=3-2$) GMC Catalog}
\label{sec31}
The GMC catalog is obtained from our CO($J=3-2$) data using the CPROPS package described in \citet{2006PASP..118..590R}.
Briefly, we search for compact emission in adjacent pairs of channels with $T_{\rm mb} > 4\,\sigma_{\rm rms}$, the so-called {\it cores}, and then for emission with $T_{\rm mb} > 2\,\sigma_{\rm rms}$ connected to such cores. 
CPROPS attempts to account for the amount of flux below the 2\,$\sigma_{\rm rms}$ cutoff by linearly extrapolating the emission profile to the $0\, {\rm K\,km\,s}^{-1}$ intensity level.
Finally, the emission identified with CPROPS must have a minimum velocity width of 2 channels ($5\,{\rm km\,s}^{-1}$) with a minimum peak of $T_{\rm mb} > 4\,\sigma_{\rm rms}$, and an equivalent size of at least the spatial resolution of 25$\arcsec$ ($\sim$100\,pc). 
Therefore, GMCs with a narrower velocity width than 2.5\,km\,s$^{-1}$ (if any) will not be identified by this analysis.

As a result, we have identified a total of 71 GMCs, whose properties are summarized in Table~\ref{p3:co32cprops}.
The columns indicate the cloud ID, the cloud position (R.A., Decl.), the velocity, 
the peak intensity, the FWHM of the velocity width ($\Delta V_{\rm FWHM}$), the major/minor axes and orientation of the GMCs,
the CO($J=3-2$) luminosity ($L^{\prime}_{\rm CO(3-2)}$), virial mass ($M_{\rm vir}$) and cross identifications with the CO($J=1-0$) and CO($J=2-1$) GMC catalogs of previous papers \citep{2007ApJ...661..830R,2012A&A...542A.108G}.
The cloud position and velocity are derived from an intensity-weighted mean over all the pixels of a GMC.
The major/minor diameters ($A_{\rm maj}, A_{\rm min}$) without beam deconvolution are also indicated.
The virial mass, calculated under the assumption that each cloud is spherical and can be parametrized by a density profile $\rho \propto r^{-\beta}$, is given by $M_{\rm vir} = 189\,\Delta V_{\rm FWHM}^2 R_{\rm deconv}\,M_{\odot}$ \citep{1987ApJ...319..730S}, where $R_{\rm deconv}$ is the deconvolved radius given by deconvolving the beam size ($\theta_{\rm beam}$) from the GMC size, i.e., $R_{\rm deconv}=\sqrt{[A^2_{\rm maj}-\theta^2_{\rm beam}]^{1/2}[A^2_{\rm min}-\theta^2_{\rm beam}]^{1/2}}$.
Note that the virial masses of 27 GMCs are not shown because their minor diameters are smaller than the beam size.

Among the 71 GMCs in our catalog, only one GMC (GMC-50) is a new identification. The 70 GMCs have been previously identified in lower-$J$ CO transition catalogs \citep{2007ApJ...661..830R,2012A&A...542A.108G}. 
GMC-50 is located outside of the observed field in \citet{2012A&A...542A.108G} and has a marginal detection in the CO($J=1-0$) integrated intensity map in \citet{2007ApJ...661..830R}, but was not identified in their catalog. 
A total of 10 GMCs in our catalog are composed of more than two smaller GMCs in previous catalogs, due to their higher resolution data \citep[$\sim$50\,pc;][]{2007ApJ...661..830R,2012A&A...542A.108G}. 
On the other hand, a few single GMCs in the \citet{2007ApJ...661..830R}'s catalog appear to be composed of more than two GMCs in our catalog. This is because of our 8--15 times better sensitivity data, since decomposition of neighboring clouds is done based on peak differences exceeding 2 $\sigma_{\rm rms}$ \citep{2006PASP..118..590R}.
The ID numbers in these cases are marked with an asterisk in Table~\ref{p3:co32cprops}. 
Since 6 GMCs (GMC-18, GMC-28, GMC-47, GMC-51, GMC-58, and GMC-71) are at the edge of the observed fields and their extents are uncertain, they are not included in further analysis and discussion.

\begin{figure}
\epsscale{1.0}
\plottwo{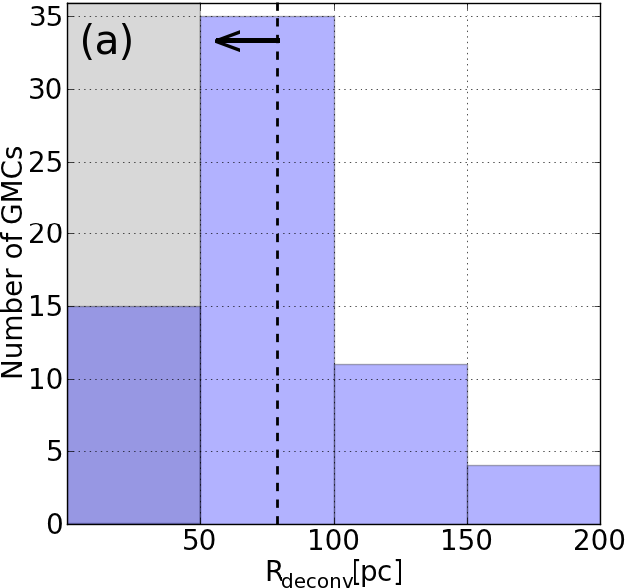}{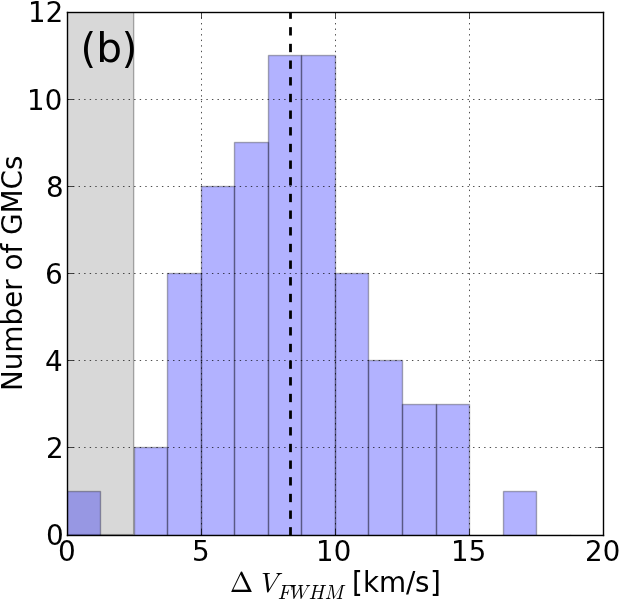}\\
\plottwo{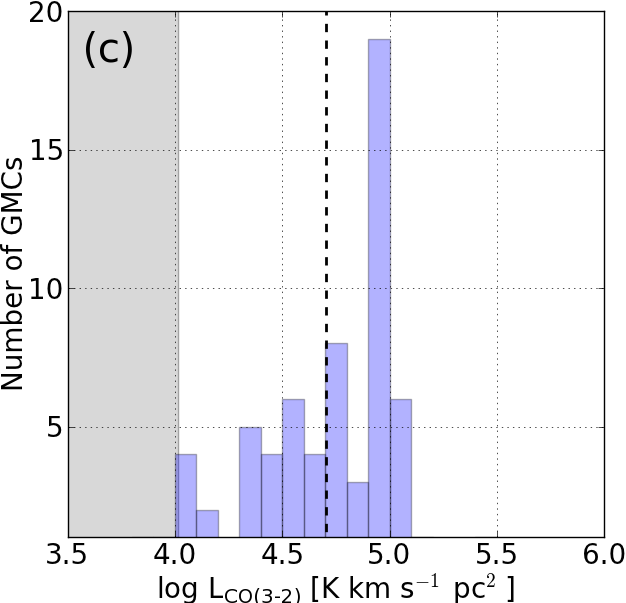}{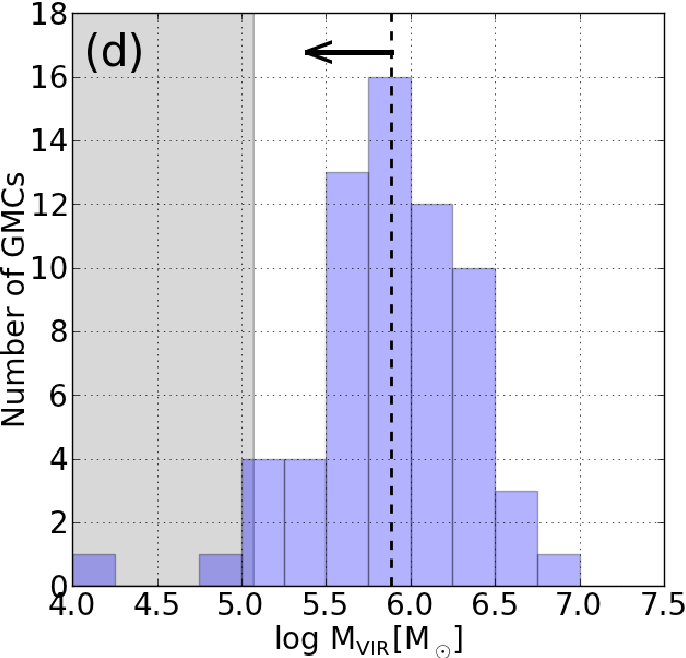}
\caption{Histograms of physical properties of GMCs as traced by CO($J=3-2$): (a) deconvolved radius ($R_{\rm deconv}$), (b) Full Width Half Maximum (FWHM) line width ($\Delta V_{\rm FWHM}$), (c) luminosity ($L^{\prime}_{\rm CO(3-2)}$), and (d) virial mass ($M_{\rm vir}$). The shaded areas indicate the detection limit. The dashed lines represent the average values: $R_{\rm deconv}\la73$\,pc, $\Delta V_{\rm FWHM} \simeq 8.3\,$km\,s$^{-1}$, $L_{\rm CO(3-2)} \simeq 5.1\times10^4$\,K\,km\,s$^{-1}$\,pc$^2$, and $M_{\rm vir}\la 7.6\times10^5$\, M$_{\odot}$.\label{p3:co32gmcprop}}
\end{figure}

Histograms of the physical properties of 65 GMCs are presented in Figure~\ref{p3:co32gmcprop}. 
The $\Delta V_{\rm FWHM}$, $R_{\rm deconv}$,
$L^{\prime}_{\rm CO(3-2)}$, and $M_{\rm vir}$ are in the range of 1.2--17.3\,km\,s$^{-1}$, 12--157\,pc, (0.8--25.8)$\times10^4$\,K\,km\,s$^{-1}$ pc$^2$, and (0.2--89.2)\,$\times$10$^5\,M_{\odot}$, respectively.
A lower limit for the virial mass is $1.2\times10^5\,M_{\odot}$, assuming an effective radius of $\sim$50\,pc and $\Delta V_{\rm FWHM} =2.5\,{\rm km\,s}^{-1}$, which are close to the instrumental limits.
Average values are $\Delta V_{\rm FWHM} \sim 8.3\,$km\,s$^{-1}$, $R_{\rm deconv}\la73$\,pc, $L_{\rm CO(3-2)} \sim 5.1\times10^4$\,K\,km\,s$^{-1}$\,pc$^2$, and $M_{\rm vir}\la 7.7\times10^5\, M_{\odot}$.
Note that we have used the non-deconvolved radius for the GMCs whose minor axes were smaller than the beam size, and thus the averaged $R_{\rm deconv}$ and $M_{\rm vir}$ should be regarded as upper limits.
These averaged values are within the typical range as traced by CO($J=1-0$) in LGGs \citep{2010ARA&A..48..547F}.

Compared to the high-resolution CO($J=3-2$) observations towards clumps in LMC,  the $R_{\rm deconv}$ and $M_{\rm vir}$ of M33 GMCs are larger than the typical values of LMC clumps, i.e. $R_{\rm deconv}=1.1$--12.4\,pc and $M_{\rm vir}=4.6\times10^3$--$2.2\times10^5\,M_{\odot}$ \citep{2008ApJS..175..485M}.
This is likely only due to our more limited spatial resolution, because the $\Delta V_{\rm FWHM}$ of M33 GMCs is similar to that of the clumps at high resolution, $\Delta V_{\rm FWHM}$ of $\sim$7\,km\,s$^{-1}$. 
This suggests that the M33 GMCs obtained in this work are composed of a single or a few small clumps at a similar velocity, but not a large amount of clumps.

\subsubsection{CO($J=3-2$) and CO($J=1-0$) distribution and $R\ratio$}
\label{sec311}
The CO($J=3-2$) and CO($J=1-0$) integrated intensity maps and also the CO intensity ratio ($R\ratio$) map for each individual GMC, with a common 25$\arcsec$ resolution, are shown in panels (a)--(c) of Figures~\ref{fig6-gmc-60} -- \ref{fig6-gmc-16} and Figures~18 -- 78 in the Appendix. 
The integrated intensity maps of CO($J=3-2$) emission have been created by using the channel maps where the emission is above 2\,$\sigma_{\rm ch}$. 
The integrated intensity maps of CO($J=1-0$) emission have been binned over the same velocity ranges as that of CO($J=3-2$) emission.

We find a general trend that the distribution of CO($J=3-2$) emission and peaks are similar to that of CO($J=1-0$)  (e.g., GMC-3 in Figure~19, GMC-5 in  Figure~21).
However, in some cases ($\sim 28\,\%$), the CO($J=3-2$) emission for some GMCs  shows more compact distribution (e.g., GMC-1 in Figure~\ref{fig6-gmc-1}, GMC-8 in  Figure~\ref{fig6-gmc-8}).
In other GMCs ($\sim 11\,\%$), the CO($J=3-2$) emission peak is offset from the CO($J=1-0$) emission peak by over half HPBW (e.g., GMC-1 in Figure~\ref{fig6-gmc-1}, GMC-7 in  Figure~23).
A further description of the spatial comparison between CO emission and massive SF sites is presented later in Section~\ref{sec3.4.1}.

We have calculated $R\ratio$ by dividing the CO($J=1-0$) map by the CO($J=3-2$) map, after masking the region where the CO($J=1-0$) emission is below 3\,$\sigma$ in its integrated intensity map.
Note that each $R\ratio$ map is masked except inside the boundary of the GMC, to avoid confusion with neighboring GMCs.

\begin{figure}
\epsscale{0.5}
\plotone{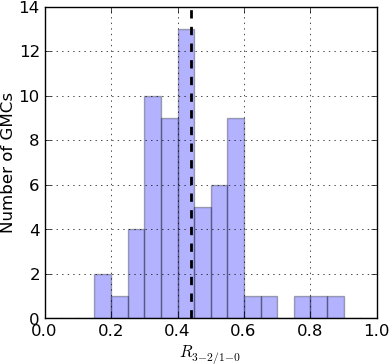}
\caption{Distribution of average $R\ratio$ over each GMC. The GMCs at the edge of the observed fields or with underivable $R\ratio$ were discarded. The dashed line represents the average value, 0.43. \label{p3:co32gmcprop_hist}}
\end{figure}

The $R\ratio$ values across GMCs vary greatly.
Some GMCs are found to have a gradient of $R\ratio$, while others have a relatively constant ratio.
The averaged $R\ratio$ is measured inside the boundary of GMCs and is provided in Column~(9) of Table~\ref{p3:co32cprops}. 
The CO($J=3-2$) emission for four GMCs (GMC-49, 56, 57 and 70) have been detected with significant signal-to-noise ratio but the CO($J=1-0$) emission has not been detected at the same position.
In case that the $R\ratio$ for these GMCs is uncertain, we calculate the lower limit of $R\ratio$ at the CO($J=3-2$) emission peak (see captions in each figure).
Figure~\ref{p3:co32gmcprop_hist} shows the histogram of $R\ratio$ for the 65 GMCs.
The averaged $R\ratio$ ranges from 0.18 to 0.89, with a mean of 0.43.
High ratios, $R\ratio$ $> 0.6$, are found at four GMCs: GMC-1, GMC-7, GMC-34 and GMC-42, which are located around the GHRs (NGC~604, NGC~595) and the vicinity of the galaxy center.

\subsection{Identification of YSGs and their Ages}
\label{p3:age}
The color magnitude diagram (CMD) is a powerful tool to provide insight into the age of the stellar component in a region of interest.
Next we describe the method to identify YSGs as well as how to estimate their ages using the CMD. The results are then used to investigate the spatial correlation between the YSGs and the GMCs identified in Section~\ref{sec31}. 

\subsubsection{Extraction of Young Stars}
\label{p3:4321}
We have created the $V$ magnitude - ($B-V$) CMD (i.e., $M_V$-($B-V)_0$ diagram) for the disk of M33 using the photometry catalog described in Section~\ref{p3:2.2}.
To obtain intrinsic absolute magnitudes and then accurate ages, the photometric measurements must be corrected for two sources of
reddening: the foreground extinction from our Galaxy and the internal reddening due to  the interstellar matter in the disk of M33.
We adopt the foreground galactic extinctions in the $B$ and $V$ bands of 0.181 and 0.139~mag, respectively \citep{1998ApJ...500..525S}.

The reddening $E(B-V)$ has been measured towards some individual clusters in M33, and ranges from 0.06 to 0.3 mag, with a typical value of 0.10 \citep{1999ApJ...517..668C,2007AJ....134.2168P}. 
Due to the different density structures of the GMCs it is necessary to consider a variable internal reddening correction. This is especially important if the stars are embedded or lie behind the GMC, as the visual extinction is then expected to be higher \citep[e.g.,][]{2010ApJ...721..686P}. 
The exact correction for a given line of sight is rather uncertain, but we have used the available $E(B-V)$ values obtained for some clusters, and the typical value of $E(B-V)=0.10$ \citep{1999ApJ...517..668C,2007AJ....134.2168P} for the remaining stellar groups.
The ratio of total to selective extinction $R_V = A_V / E(B-V) \simeq 3.1$  in our Galaxy \citep[][]{1979ARA&A..17...73S} is used to obtain the visual extinction $A_V$.

We have used the theoretical isochrones in the Padova stellar population synthesis models \citep[][and references therein]{2008A&A...482..883M,2010ApJ...724.1030G} to estimate the age of the stars.
The adopted distance to M33 \citep[840\,kpc;][]{2001ApJ...553...47F} corresponds to a distance modulus  ($m-M)_0 = 24.6$.
Note that the metallicity ($Z$) varies from $0.6\,Z_{\odot}$ in the central region to 0.4\,$Z_{\odot}$ at 5\,kpc galaxy radius \citep{2011ApJ...730..129B}.

\begin{figure}
  \begin{center}
  \epsscale{1.0}
    \plotone{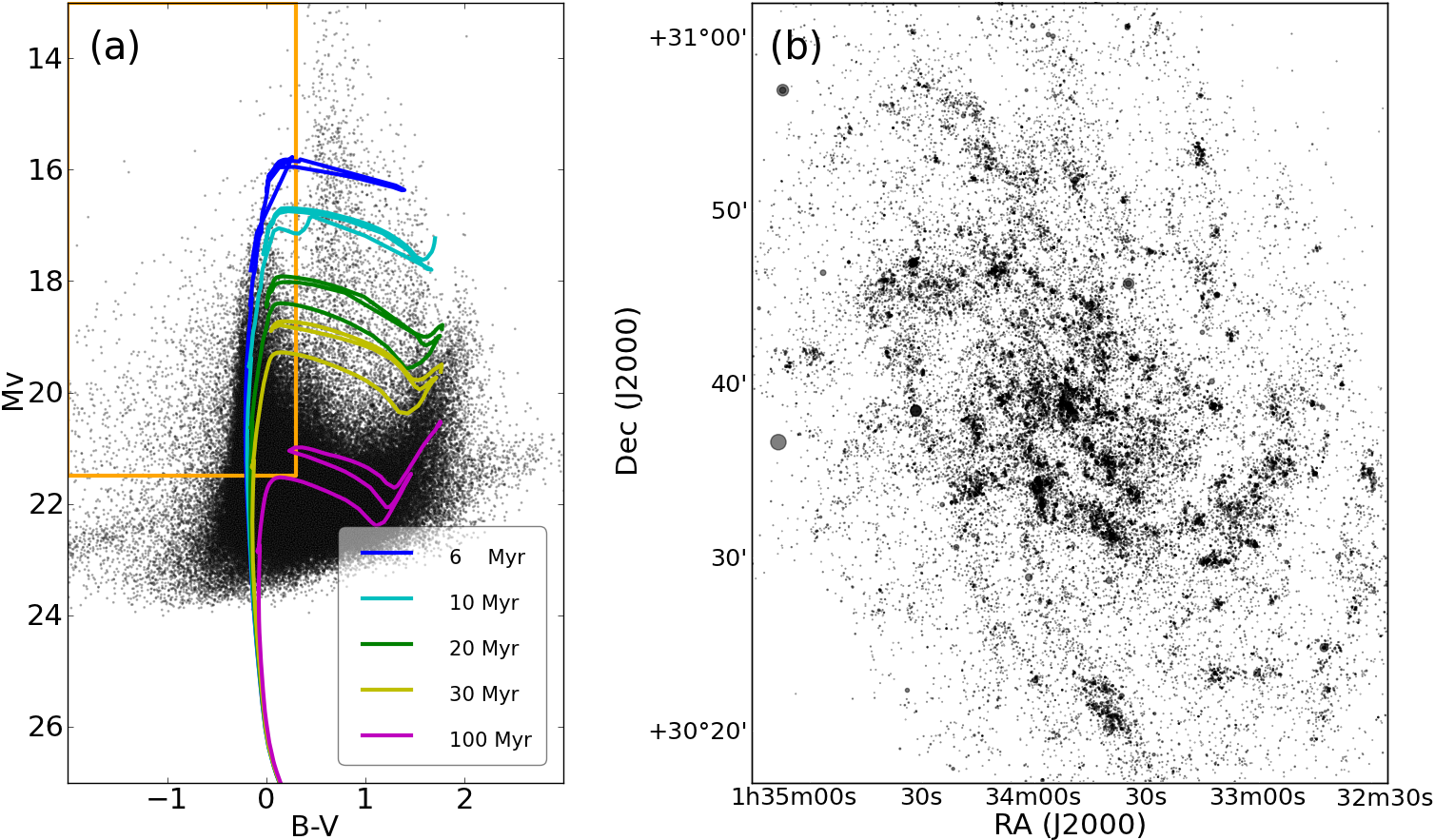}
  \end{center}
  \caption{{\bf (a)} The Color-Magnitude Diagram of the entire disk of M33. 
The five lines represent isochrone tracks of 6\,Myr (blue), 10\,Myr (cyan), 20\,Myr (green), 30\,Myr (yellow) and 100\,Myr (pink). A metallicity of 0.5 Z$_\odot$ was used. The orange boundary (horizontal and vertical lines) indicates our color and magnitude selection criteria, $ (B-V)_0 < 0.3 $ and $ M_V < 21.5 $.
{\bf (b)} The spatial distribution of identified young stars, selected using the same color and magnitude criteria. Note that the point sizes in this plot are proportional to the stellar brightness in $V$-band. \label{p3:fig3e}}
\end{figure}

Figure~\ref{p3:fig3e}(a) shows a $M_V$-($B-V)_0$ diagram for the stars in M33 with five stellar isochrones from the Padova models \citep[][and references therein]{2008A&A...482..883M,2010ApJ...724.1030G}, assuming $Z=0.5\,Z_{\odot}$,  ages of 6, 10, 20, 30 and 100\,Myr, and an average internal extinction correction of $A_V \sim 0.22$ \citep{2007AJ....133.2393M,2007MNRAS.379.1302B}. 
\citet{2006AJ....131.2478M,2007AJ....133.2393M} have estimated that $\sim 40$\,\% of stars have $(B-V)_0$ between $0.3-1.0$, but $M_V$ in the range of 14.6--19.6 are likely foreground objects \citep{2007MNRAS.379.1302B}. Taking into account photometric errors and foreground galactic extinction, the criterion of $(B-V)_0 < 0.3$ is chosen to avoid the foreground stars.
The faint limit of the $V$-band magnitude of $M_V < 21.5$ is set to select only young massive stars ($<\,100$\,Myr). 
The minimum (initial) stellar masses at $M_V\sim21.5$ for ages of 6, 10, 20, 30, 100\,Myr correspond to 14, 13, 10, 8, and $5\,M_{\odot}$, respectively.
The orange boundary in Figure~\ref{p3:fig3e}(a) indicates our selection criteria, $ (B-V)_0 < 0.3 $ and $ M_V < 21.5 $, an area that should represent young stars.

The selected young stars are plotted in Figure~\ref{p3:fig3e}(b), where the point size in the plot is proportional to the stellar brightness in $V$-band.
These young stars ($<$ 100\,Myr) are found to be distributed in a structure with two main arms and several weak multiple arms, which is in good agreement with \citet{1985Ap&SS.116..341I}.

\subsubsection{Young Stellar  Number Density Map, Identification of YSGs  and their Ages}
\label{p3:5322}
We describe the method to search for YSGs and how to estimate their ages.
The term ``YSGs''  refers to the concentration of young stars selected using the criteria in Section~\ref{p3:4321}. These will include both (young) star clusters as well as OB associations. Generally, a star cluster is defined to be a gravitationally bound system of several stars or more, whose concentration is larger than that of the surrounding stellar background, with a typical size of a few parsec to tens of parsecs \citep[e.g.,][and references therein]{2008AJ....135.1567H}.
The definition of an OB association is usually a single, loosely bound or unbound concentration of early-type luminous stars, typically extending several tens to over hundred parsecs \citep[e.g.][and references therein]{2000AJ....119.1737G}.
The typical masses of young ($<10$\,Myr) clusters and associations are 10--10$^4\,M_{\odot}$ and 10$^3$--10$^6\,M_{\odot}$, respectively \citep[e.g.,][]{2003AJ....126.1836H,2010AJ....140..379K}.
Here we do not aim to
distinguish between them because our interest is to investigate the collective properties of YSGs associated with GMCs.

In previous studies, clusters or associations have been selected to be as stellar groups just by visual inspection, and therefore the selection may have been subjective.
Recently, a more objective approach has been carried out by \citet{2000AJ....119.1737G,2010ApJ...725.1717G}, who have performed a stellar cluster identification in NGC\,6822 and LMC using the stellar surface density map, and have shown the hierarchical structure of blue stellar clusters.

\begin{figure}
  \begin{center}
    \plotone{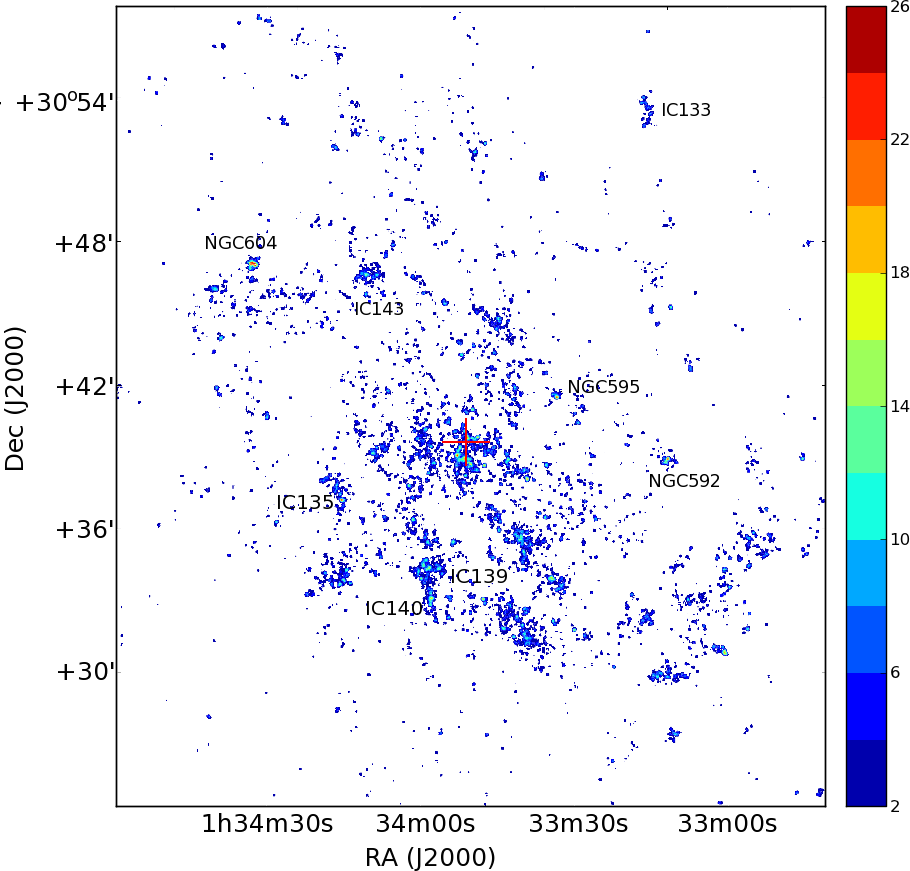}
  \end{center}
  \caption{The number density ($n^{\ast}$) map of young stars shown in Figure~\ref{p3:fig3e}(b), measured in $30\,{\rm pc}\times 30\,{\rm pc}$ area. The color range spans from 2 to 26 stars per area, in 2 stars per area color steps. This corresponds to 0.002 -- 0.031 stars per pc$^2$, in 0.002 stars per pc$^2$ steps. The number density of $n^{\ast} < 2$ stars per area is not shown. Prominent H{\sc ii} regions are also labeled for reference. A cross symbol in the middle of the image represents the galaxy center.\label{p3:fig3d}}
\end{figure}

To select YSGs in an objective manner for M33 as well, we define them as an excess in the number of young stars per unit of area ($n^{\ast}$) following \citet{2000AJ....119.1737G,2010ApJ...725.1717G}'s approach. 
We set the unit of area to the same scale as the pixel size of CO($J=3-2$) map, a $8\arcsec\times8\arcsec$ box (corresponding to $\sim30\,{\rm pc}\times 30\,{\rm pc}$), which is comparable to the typical GMC size.
The pixel size of the stellar number density map is set to half of the pixel size of the CO($J=3-2$) map, 15\,pc.
Figure~\ref{p3:fig3d} shows the number density map of young stars per each $30\,{\rm pc}\times 30\,{\rm pc}$ area in M33 disk.
It becomes more clear than in Figure~\ref{p3:fig3e}(b) that most of the regions with $n^{\ast} > 2$ stars per unit area are distributed along the spiral arms of M33. 
The maximum value of stellar number density in M33 is found in the center of NGC~604,  $n^{\ast} = 28$ stars per area.
Over-density regions ($n^{\ast} > 15$ stars per area) coincide with prominent H{\sc ii} regions such as NGC~604, NGC~595, NGC~592, IC~135, IC~139, IC~140, and the galaxy center, as labeled in Figure~\ref{p3:fig3d}.

\begin{figure}
  \begin{center}
    \plotone{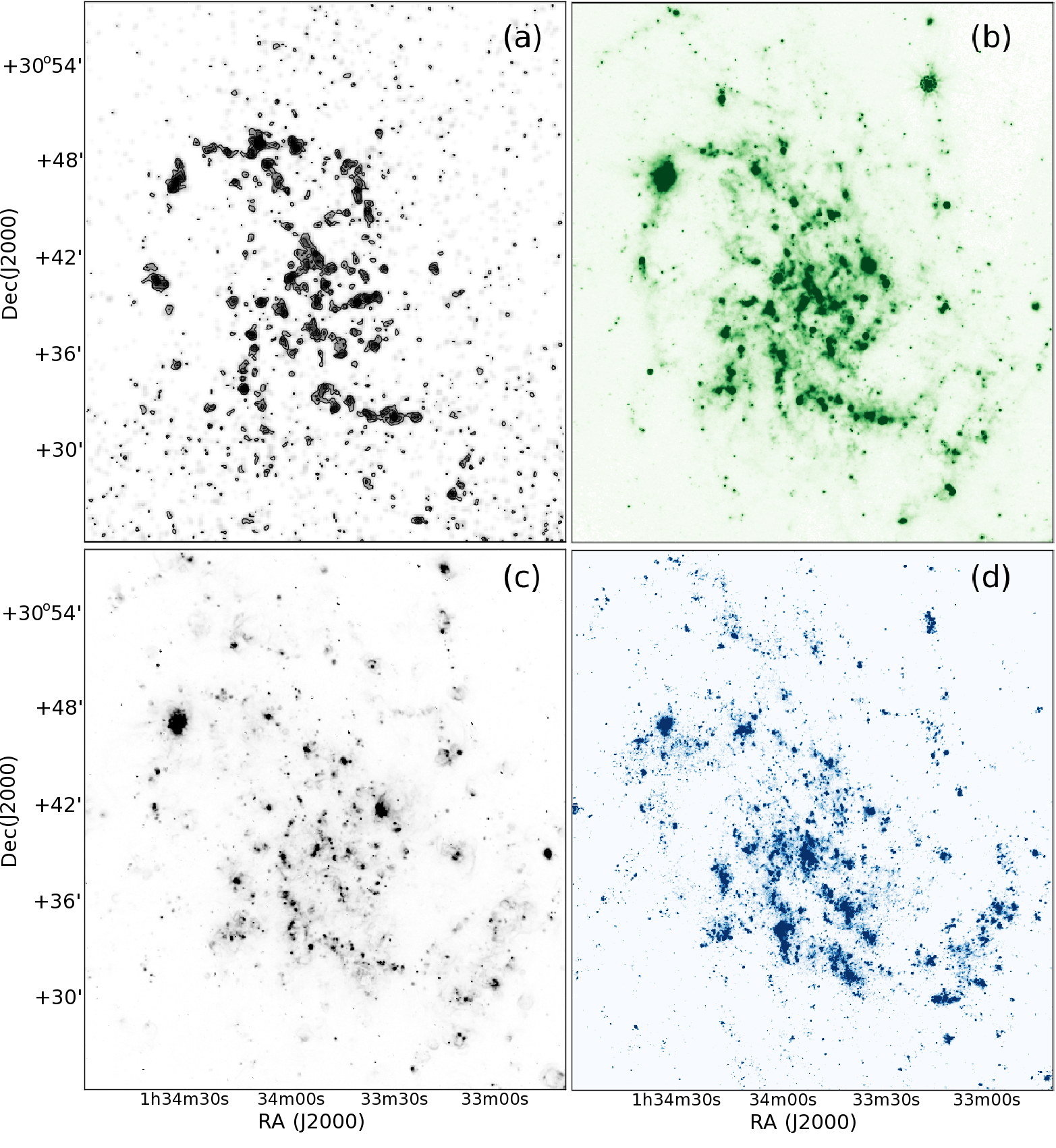}
  \end{center}
  \caption{{\bf (a)} CO($1-0$) integrated intensity map (see Section~\ref{p3:sec322}). Contour levels are 1, 2, 3,..., 8\,$\sigma$, where the noise level 1\,$\sigma$ = 1.7\,K\,km\,s$^{-1}$. Images of {\bf (b)} 24\,$\micron$, {\bf (c)} H$\alpha$, and {\bf (d)} FUV (See Sections~\ref{p3:2.2} and ~\ref{p3:5322}.) \label{p3:fig4}}
\end{figure}

Next we compare the number density map with other wavelength data.
Figure~\ref{p3:fig4} shows the M33 disk in CO($J=1-0$), $24\,\micron$, H$\alpha$, and FUV.
The background-subtracted FUV map is a reproduction from \citet{2010A&A...510A..64V}. 
Interestingly, the distributions of the molecular gas traced by CO($J=1-0$) in Figure~\ref{p3:fig4}(a) and the young stars in Figure~\ref{p3:fig3d} exhibit slightly different patterns between the northern and the southern sides of the disk: molecular gas is more abundant in the northern side of the disk, while the opposite is true for young stars.
Note that this asymmetrical pattern of the young stellar distributions is not because the photometric depth is inhomogeneous over the disk, as the completeness limits are similar (below $V\sim21.5$\,mag) both in the northern and southern disk.

The distributions of $24\,\micron$ and H$\alpha$ emissions show multiple spiral arms similarly to the density map of young stars, but their spatial distribution patterns are not always similar:  the $24\,\micron$ and H$\alpha$ emissions are enhanced at GHRs such as NGC~604 and NGC~595 compared to other disk regions, while the surface density map of young stars is more pronounced along the spiral arms (Figure~\ref{p3:fig4}b, c).
The distribution of the young star number density is remarkably similar to that of the FUV emission in Figure~\ref{p3:fig4}(d), which traces the spiral structure and reproduces the north-south asymmetry.
This is consistent with FUV being a diagnostic of the younger (30--100\,Myr) stellar population \citep{2009A&A...493..453V,2010A&A...510A..64V}.

In this work, we adopt 5 young stars per $30\,{\rm pc}\times 30\,{\rm pc}$ area as the minimum surface stellar number required to be classified as a YSG, and then we estimate its age from the CMD, assuming that stars are coeval.
First,  we search a region with a peak of $n^{\ast} > 5$ stars per area around each GMC, then determine the extent of the YSG at a level of $n^{\ast} = 2$ stars per unit area.
A Gaussian is then fitted to the radial profile of the young stellar distribution down to $n^{\ast} > 2$, and the extent of YSGs is set as the FWHM of that Gaussian fit assuming a circular shape ($r_{\rm cl}$).
Second, we have created the $M_V$-($B-V)_0$ diagrams for the stars contained within the radius of $r_{\rm cl}$.
Finally we estimate their age using the Padova models.

\begin{figure}
  \begin{center}
    \plotone{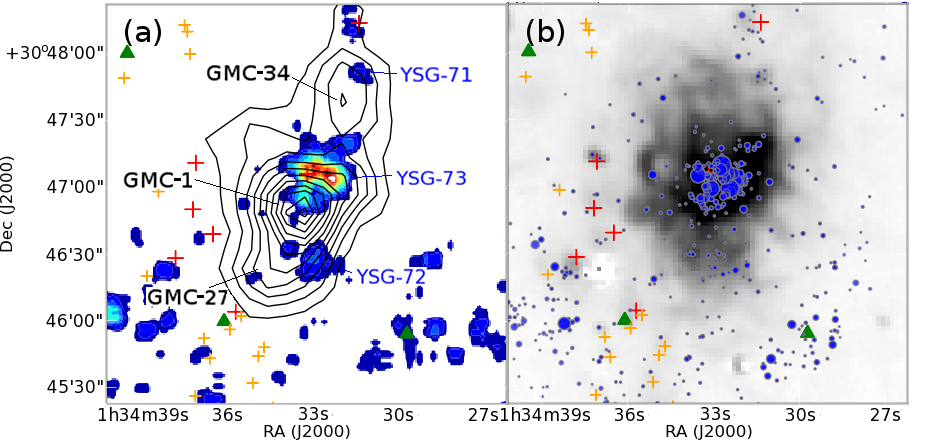}
  \end{center}
  \caption{{\bf (a)} The stellar density map toward NGC~604, overlaid with CO($J=3-2$) contours. Contour levels are 1, 2, 3, ... and 13\,K\,km\,s$^{-1}$. YSG-XX and GMC-XX are the IDs from our young stellar group and CO($J=3-2$) GMC catalogs. 
The crosses are the H{\sc ii} regions from \citet{1987A&A...174...28C} (red), and \citet{1999PASP..111..685H} and \citet{1997PASP..109..927W} (orange). The green triangles are the 24\,$\micron$ sources in \citet{2007A&A...476.1161V}. The color range is the same as in Figure \ref{p3:fig3d}. 
{\bf (b)} The distribution of individual young stars, overlaid on the gray scale image of H$\alpha$ emission of the NGC~604 region. The blue circles represent young stars which are selected with our color and magnitude criteria (see Figure~\ref{p3:fig3e}) and their sizes are proportional to the stellar brightness in $V$-band. Note that only the stars inside the radius of $90\arcsec$ from the center of the NGC~604 are shown. \label{p3:fig3a}}
\end{figure}

As an example, Figure \ref{p3:fig3a}(a) shows the number density map of young stars, overlaid on the CO($J=3-2$) map of the NGC~604 region.
Three GMCs are identified, GMC-34, GMC-1 and GMC-27, from north to south in this region.
We find three peaks of $n^{\ast} > 5$ stars per area around these GMCs, at position ($\alpha_{J2000}$, $\delta_{J2000}$) = ($1\h34\m31\fs4$, 
 $30\arcdeg47\arcmin48\farcs0$), ($1\h34\m33\fs4$, 
 $30\arcdeg47\arcmin03\farcs5$) and ($1\h34\m33\fs0$, $30\arcdeg46\arcmin23\farcs8$) from north to south, which we refer to as YSG-71, YSG-73 and YSG-72, respectively.
The radii of these YSGs are measured to be $r_{\rm cl}=9.5\arcsec$ (39\,pc),  $21.2\arcsec$ (86\,pc) and $10.4\arcsec$ (42\,pc), respectively.

\begin{figure}
  \begin{center}
    \plotone{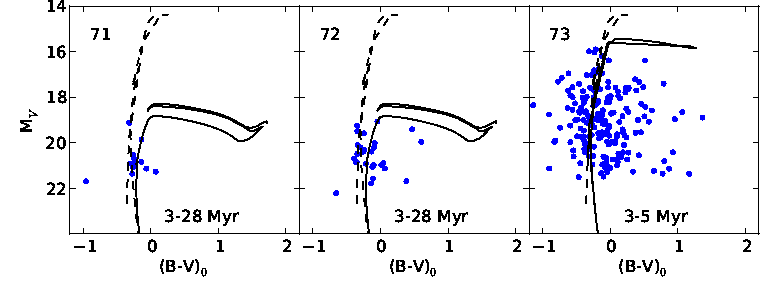}
  \end{center}
  \caption{The color-magnitude diagrams of the three identified YSGs around NGC~604: YSG-71, YSG-72 and YSG-73 (NGC~604), from left to right panel.
The dots denote the stars within the selection radius, $9\farcs5$ (39\,pc), $10\farcs4$ (42\,pc) and  $21\farcs2$ (86\,pc), respectively, which correspond to the extent at a $n^{\ast} = 2$ stars per area level. 
The solid and dashed lines are the Padova theoretical isochrone tracks of the upper and lower limit of the estimated age, respectively. The identification number in the catalog and the derived age of each YSG are shown at the upper left and lower right corners of each panel, respectively.\label{p3:fig3b}}
\end{figure}

Next, we derive ages for each YSG by fitting theoretical isochrones to the CMDs of resolved stars.
Figure~\ref{p3:fig3b} shows the CMD of the three identified YSGs above, YSG-71, YSG-72 and YSG-73.
Dot symbols represent the stars within the radius of $r_{\rm cl}$ and the lines represent the Padova theoretical isochrone tracks, assuming a metallicity of 0.5\,$Z_{\odot}$.
Solid lines represent the upper limit of the estimated ages, regarding the magnitude and color of the brightest main-sequence (MS) star as MS turnoffs for a young star group.
In this case, the more massive stars are assumed to have already evolved from the MS lines at ($B-V)_0\sim0$ and less massive stars are still in the MS. 
Dashed lines represent the lower limit of the estimated ages, considering that all stars in the YSGs are still in the MS at ($B-V)_0\sim0$, without any more massive stars in the group evolving to red giant branches. 
Although chances are low that these massive stars are detected at redder colors, because the more massive the stars, the faster they evolve, no detection of stars at ($B-V)_0 > 0$ can suggest that more massive stars have not existed before.
Note that in either case, we have used only stars at ($B-V)_0 < 0.3$ to avoid foreground stars (Section~\ref{p3:4321}).
The measurable age limit of these YSGs using the Padova models is 3\,Myr.
In this way, the estimated ages of the three YSGs YSG-71, YSG-72 and YSG-73 are in the range of 3--28\,Myr, 3--28\,Myr and 3--5\,Myr, respectively.
The first two YSGs include a smaller number of stars (11 and 21 O stars for YSG-71 and YSG-72, respectively) and mostly less bright stars ($M_V\ga19$\,mag), which makes a larger uncertainty in the estimations. 

Figure~\ref{p3:fig3a}(b) shows the distribution of the young stars, overlaid on the H$\alpha$ map (grayscale). 
The size of the blue circle representing each star is proportional to the brightness in $V$-band, that is, the larger, the brighter.
Many bright young stars are found in YSG-73 and concentrated in the center of the H{\sc ii} region. In fact, NGC~604 is known to be the most luminous and massive H{\sc ii} region in M33, containing more than 200 O-type stars \citep{1993AJ....105.1400D,1996ApJ...456..174H, 2000MNRAS.317...64G, 2003AJ....125.3082B}.
The other two YSGs (YSG-71 and YSG-72) do not have cross identifications in previous cluster catalogs, but are included in previous star complex and OB association catalogs \citep[No. 46 in ][]{2005PASRB...4...75I, 1980ApJS...44..319H}. 

\subsubsection{The Catalog of YSGs}
We have identified 75 YSGs in M33 and have estimated their ages as described in Section~\ref{p3:5322}. 
Note that only YSGs associated with the identified GMCs are included (see Section~\ref{p3:association}).

\begin{figure}
  \begin{center}
    \plotone{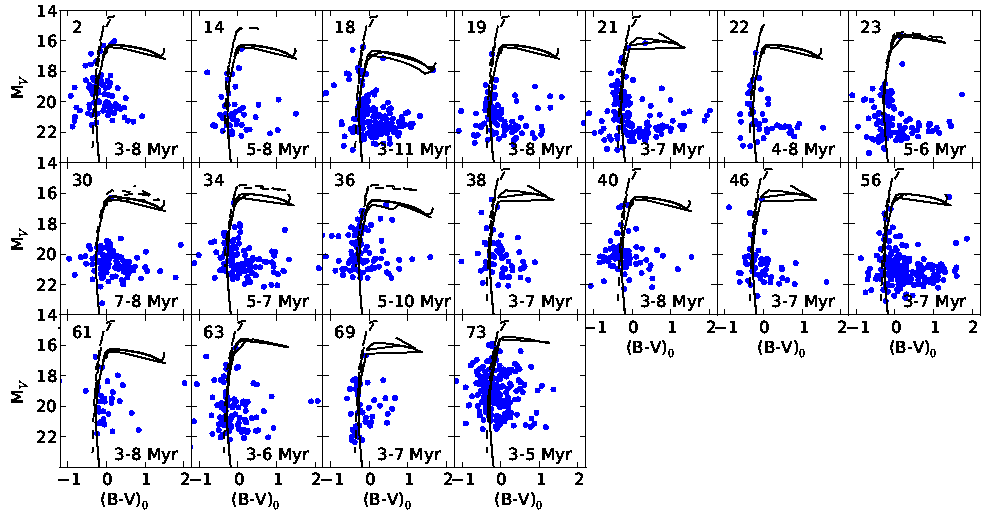}
  \end{center}
  \caption{The color-magnitude diagrams of 18 YSGs (among the 75 in the catalog), whose upper limit ages are estimated to be less than 10\,Myr. Dots denote the stars within the selection extent ($r_{\rm cl}$) in Table~\ref{p3:ys}.
The lines and labels are the same as in Figure~\ref{p3:fig3b}.
\label{p3:cmd1}}
\end{figure}

\begin{figure}
  \begin{center}
    \plotone{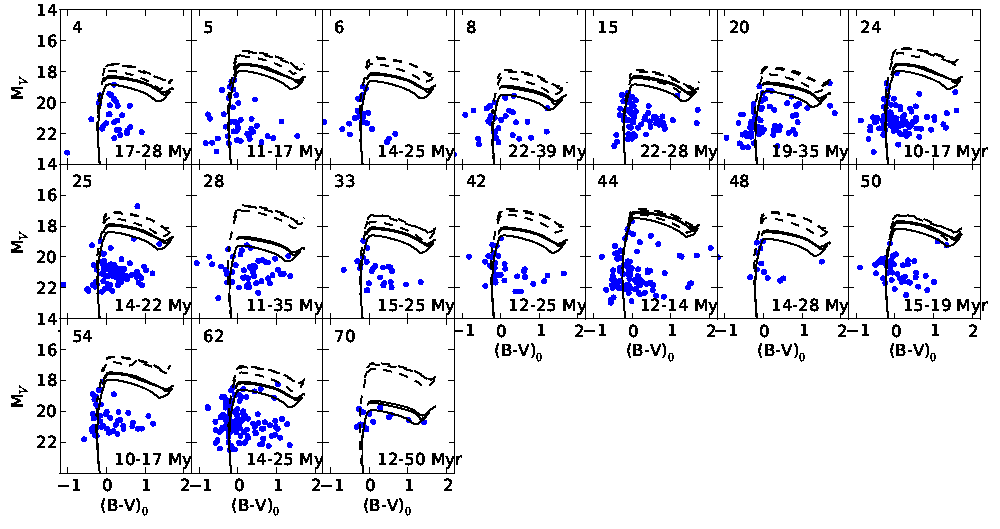}
  \end{center}
  \caption{Same as Figure~\ref{p3:cmd1} but for 17 YSGs (among 75 in the catalog), whose ages are estimated to be older than 10\,Myr. \label{p3:cmd2}}
\end{figure}

\begin{figure}
  \begin{center}
    \plotone{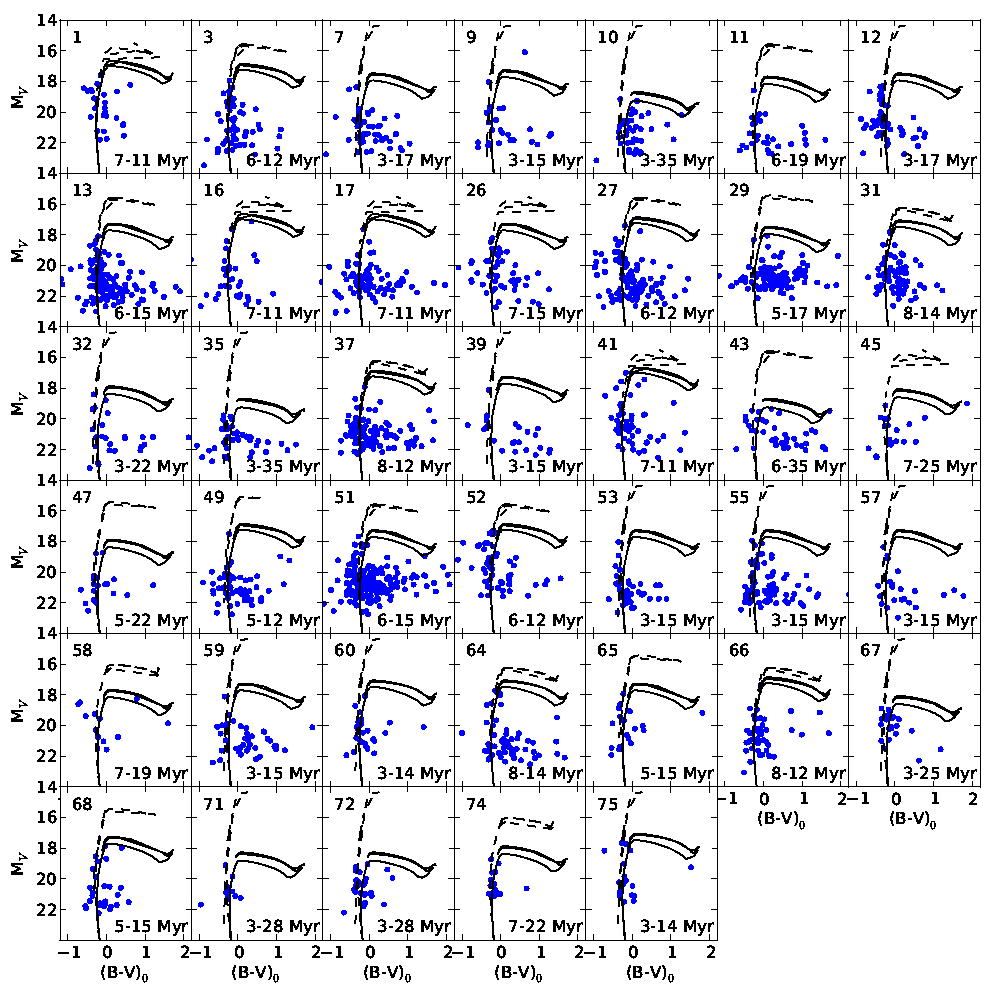}
  \end{center}
  \caption{Same as Figure~\ref{p3:cmd1} but for the remaining 40 YSGs not included in Figures~\ref{p3:cmd1} and \ref{p3:cmd2}. A total of 25 of these are estimated to be younger than 15\,Myr.  \label{p3:cmd3}}
\end{figure}

Figures~\ref{p3:cmd1}--\ref{p3:cmd3} show the CMDs of 75 YSGs with the adopted Padova theoretical isochrone tracks. Similarly to Figure~\ref{p3:fig3b}, solid and dashed lines represent the upper and lower limits of the estimated ages. Blue dots represent the stars within the $r_{\rm cl}$ of the YSGs.
In Figures~\ref{p3:cmd1}, \ref{p3:cmd2}, and \ref{p3:cmd3}, we show the CMDs of 18 YSGs younger than at most 10\,Myr, 17 YSGs older than at least 10\,Myr, and the remaining 40 YSGs, respectively.

Table~\ref{p3:ys} shows the properties of the YSGs identified in our catalog, including positions, size ($r_{\rm cl}$), number of O stars, $V$-magnitude of the brightest O star in the YSG, applied reddening correction, estimated age and cross identifications.
The number of O stars are counted only for those within our criterion on the $M_V$-($B-V)_0$ diagrams, i.e., $(B-V)_0 < 0.3$ and $M_V < 21.5$.
Note that this stellar age provides just an upper limit for the YSGs, partly because reddening correction is uncertain.
Twenty-three YSGs do not coincide with previously identified clusters, but are included in star complex and OB association catalogs \citep{2005PASRB...4...75I,1980ApJS...44..319H}.
The remaining are found to have cross identifications with previous cluster catalogs \citep[SM;][]{2007AJ....134.2168P,2010ApJ...720.1674S}.

The radii of the YSGs range from $6\farcs6$ to $22\farcs4$ (27--91\,pc) and the average is 11$\arcsec$ (46\,pc).
The number of O stars in a YSG ranges from 9 to 169 stars.
The total stellar mass of the YSGs are estimated to be $10^{3.5}-10^{4.7}\,M_{\odot}$, assuming that the YSGs are characterized by a Salpeter IMF (slope -2.35), with masses spanning 0.1--58\,$M_{\odot}$.
These sizes and masses are comparable to the typical values of OB associations  \citep[e.g.,][]{1995ApJS..101...41B,1996A&A...314...51B, 2010AJ....140..379K}. 
It is possible to estimate the stellar spectral type of the primary ionizing star contained in a YSG from the brightest $V$-magnitude and $(B-V)_0$ color.
The most massive stars are found in YSG-21 and YSG-73 and their masses are estimated to be 51\,$M_{\odot}$, corresponding to O3.5V \citep{2005A&A...436.1049M}.
The brightest stars in YSG-39 and YSG-60 are less massive than in any other YSGs, 8\,$M_{\odot}$ (corresponding to B2V).
The latter is comparable to small clusters in our Galaxy and LMC  that can be ionized by a single mid-O or B0 star \citep[e.g.,][]{1973A&A....24..219P,1994AJ....108.1674W}.
The ages of the YSGs are estimated to be in the range between $4\,\pm\,1$\,Myr and $31\,\pm\,19$\,Myr, with an average of $12\,\pm\,5$\,Myr.

We can compare the ages with other catalogs.
Although the YSGs in our catalog do not fully overlap with the clusters whose ages have been previously determined (SM; PPL; SSGH), only 5 YSGs coincide, YSG-1, YSG-13, YSG-41, YSG-63, and YSG-67.
The previously derived ages are 25$^{+38}_{-14}$\,Myr (PPL), 13\,Myr (SM), 16$^{+4}_{-3}$\,Myr (PPL) or 21\,Myr (SM), 5\,Myr (SM) or 13$^{+7}_{-5}$\,Myr (PPL), and 40$^{+60}_{-24}$\,Myr (PPL), respectively, while they are $9\,\pm\,2$\,Myr, $11\,\pm\,5$\,Myr, $9\,\pm\,2$\,Myr, $5\,\pm\,2$\,Myr and $14\,\pm\,11$\,Myr in our estimation.
Compared to the results for YSG-1, YSG-41, YSG-63 and YSG-67 in PPL who have used a similar method, we notice that our estimations of ages are slightly younger than them but consistent within the uncertainties.
This slight difference is mostly because they have used a small member star selection radius (within 1--2$\arcsec$) of a compact stellar cluster and also because they have applied different theoretical isochrones in the Padova models, corresponding to $Z=0.2\,Z_{\odot}$. 
Besides, even though a cluster in their catalogs is associated with one of our YSGs, in general it only represents a small portion of the YSG, and thus it is not clear in principle whether they represent exactly the same entity.
For the ages of the other two YSGs (clusters) in known GHRs (NGC~595 and NGC~604), YSG-2 and YSG-73, they had been previously estimated to be 4--6\,Myr \citep{1990ApJ...364..496D,1996AJ....111.1128M} and 3--5\,Myr, respectively, using optical spectroscopy combined with an instantaneous starburst model \citep[e.g.,][]{2000MNRAS.317...64G}.
The ages of these young clusters are in good agreement with our results, $6\,\pm\,3$\,Myr and $4\,\pm\,1$\,Myr, respectively.

\subsection{H{\sc ii} Regions and YSGs Associated with GMCs}
\label{p3:association}
In this section, we investigate the associations of H$\alpha$, 24\,$\micron$ sources, and the identified YSGs, with the GMCs. 
These associations are listed in Table~\ref{p3:gmctype}.
H{\sc ii} regions are tracers of the younger stellar population ($<$ 10\,Myr), while 24\,$\micron$ sources represent massive SF sites embedded in dust.
When the extent of H{\sc ii} regions, 24\,$\micron$ sources and YSGs are confined within the boundary of a GMC, they are treated as being associated with the GMC.

The H{\sc ii} region catalogs \citep{1974A&A....37...33B, 1987A&A...174...28C,1999PASP..111..685H} include those H{\sc ii} regions with H$\alpha$ luminosities larger than $10^{34.4}$\,erg\,s$^{-1}$ \citep{1997PASP..109..927W}.
Therefore, the sensitivity of the survey is high enough to detect the equivalent to a single O9V type star \citep{2005A&A...436.1049M}, assuming a standard relation between ionizing photon rates and H$\alpha$ luminosities \citep[][and references therein]{1996AJ....111.1252M}. 
For the 24\,$\micron$ sources, the faint end of the 24\,$\micron$ luminosity in the catalog \citep{2007A&A...476.1161V} is $\sim10^{37.7}$\,erg\,s$^{-1}$, corresponding just to an H{\sc ii} region illuminated by a single B1.5V star.
Note that the 24\,$\micron$ source catalog includes discrete sources such as H{\sc ii} regions, supernovae remnants and planetary nebulae, but 
most of the 24\,$\micron$ sources are classified as H{\sc ii} regions \citep{2007A&A...476.1161V}, which contain embedded young stellar clusters with ages of 3--10\,Myr \citep{2011A&A...534A..96S}.

The 24\,$\micron$ sources have generally corresponding H{$\alpha$} emitting sources, but there are some cases of 24\,$\micron$ sources without corresponding H{\sc ii} source (see the last column in Table~\ref{p3:gmctype}).
All GMCs except GMC-60 are found to be associated with H{\sc ii} regions and/or 24\,$\micron$ sources, while 51 GMCs are found to be associated with YSGs.
 In 10 GMCs no H{\sc ii} region exists but 24\,$\micron$ sources have been identified, while in 16 GMCs there are H{\sc ii} regions without corresponding 24\,$\micron$ sources.

The H$\alpha$ and 24\,$\micron$ luminosities, $L({\rm H}\alpha)$ and $L(24\,\micron)$, and SFR for each GMC are also listed in Table~\ref{p3:gmctype}, where the SFR for each GMC is calculated from the $L({\rm H}\alpha)$ and $L(24\,\micron)$ using the relation between the extinction-corrected H$\alpha$ emission and the SFR \citep{2007ApJ...666..870C}.
The $L({\rm H}\alpha)$ and $L(24\,\micron)$ for each GMC, are measured with a circular aperture of size equal to that of the associated GMC (100 to 400\,pc diameter).
These aperture photometry measurements include local background subtraction.
Determination of the local background is done by fitting the emission over the aperture with Gaussian functions.
These aperture sizes are similar or larger than a typical H{\sc ii} region (1\,pc -- 100\,pc) and thus they may potentially contain several H{\sc ii} regions.
A more careful local background subtraction for these H{\sc ii} regions would just result in even larger differences in luminosities between a small H{\sc ii} region and a large one because the local background contamination tends to be systematically large at the faint end \citep{2011ApJ...735...63L}.
The $L({\rm H}\alpha)$ and  $L(24\,\micron)$ range from $2.49\times10^{35}$ to $7.67\times10^{39}$\,erg\,s$^{-1}$, and from $1.67\times10^{37}$ to $4.04\times10^{40}$\,erg\,s$^{-1}$, respectively. The most luminous source both in H$\alpha$ and 24\,$\micron$ emission is found at GMC-1, corresponding to NGC~604.
The SFR of the M33 GMCs is characterized by a large scatter, ranging from $8.89\times10^{-5}$ to 0.216\,$M_{\odot}\,{\rm yr}^{-1}$.

We estimate the averaged H$\alpha$ attenuation over a GMC (i.e., a few 100 pc scale) using $L(24\,\micron)/L(\rm H\alpha)$ ratio as a proxy \citep[][]{2007ApJ...668..182P,2007ApJ...671..333K}. 
The H$\alpha$ attenuation is given by $A_{\rm H\alpha}=2.5\, {\rm log} [1+0.038\,L(24\,\micron)/L({\rm H\alpha})]$ \citep{2005ApJ...633..871C,2007ApJ...668..182P}. We list  $A_{\rm H\alpha}$ in Table~\ref{p3:gmctype}. 
The $A_{\rm H\alpha}$ is generally low (less than 1\,mag) and in good agreement with previous studies \citep{2010A&A...521A..41G,2011A&A...534A..96S}.
The averaged $A_{\rm H\alpha}$ is 0.4\,mag ($A_V\sim0.5$) and is consistent with the adopted extinction for individual YSGs in Section \ref{p3:4321}.
Relatively highly obscured sources ($A_{\rm H\alpha} >$ 2\,mag) are found in three GMCs: GMC-6, 15 and 68. The ages for the YSGs associated with these GMCs may be underestimated. Note that highly obscured regions ($A_{\rm H\alpha} >$ 3\,mag) have been found in other nuclei of galaxies \citep[][]{2007ApJ...668..182P}, but we can discard that possibility in M33.

\subsection{GMC Types}
\label{p3:sec534}
We classify the GMCs into four types according to the age of the associated YSGs and H{\sc ii} regions:\\
\indent Type~A: GMCs are not associated with H{\sc ii} regions nor YSGs;\\
\indent Type~B: GMCs are associated with H{\sc ii} regions, but not with any YSG;\\
\indent Type~C: GMCs are associated with H{\sc ii} regions and young ($<10$\,Myr) stellar groups;\\
\indent Type~D: GMCs are associated with H{\sc ii} regions and relatively old (10--30\,Myr) stellar groups.

Note that averaged values between the upper and lower limits of the stellar ages are used for this classification.
If a GMC is associated with several YSGs of different ages, the classification is done based on the youngest stellar group.

Table~\ref{p3:gmctype} lists the classification of each M33 GMC into one of these four types.
Out of  the 65 GMCs, 1 (1\,\%), 13 (20\,\%), 29 (45\,\%), and 22 (34\,\%) are found to be Types~A, B, C, and D, respectively.
Table~\ref{p3:class} summarizes these statistics.
The Type~C GMCs are the majority among all GMC types.

The estimated age of some YSGs may be uncertain.
We select 44 GMCs with accurate age estimations for their associated YSGs in order to check how robust the relative percentage is among the different types. In this regard, only YSGs whose age upper limit is smaller than 10\,Myr (see Figure~\ref{p3:cmd1}) and lower limit larger than 10\,Myr (see Figure~\ref{p3:cmd2}) are used.
With this condition we obtain that 1 (2\,\%), 13 (30\,\%), 19 (43\,\%), and 11 (25\,\%) are classified as Types~A, B, C, and D, respectively.
The 44 selected GMCs are indicated in Table~\ref{p3:gmctype}.
We confirm the trend in the relative percentages, being Type~C GMCs the majority among all GMC types.

A caveat in the Type~B and C classifications is that the percentages may be partly influenced by the instrumental sensitivity in \citet{2006AJ....131.2478M}'s survey, which may limit the detection of some YSGs. Given the instrumental sensitivity of \citet{2006AJ....131.2478M}'s survey, some of Type B sources possibly contain YSGs that have not been detected. However, the sensitivity restricts the contained YSGs to be less massive than $13\,M_{\sun}$ ($\sim$B1V star) (Section \ref{p3:4321}), and Type C (YSGs with ages $< 10$\,Myr) are typically more massive than that.

The three GMC classifications in the LMC, Types~{\sc i}, {\sc ii} and {\sc iii} \citep{2009ApJS..184....1K} correspond to Types~A, B and C in our classification, respectively (see also Table~\ref{p3:class}). In the LMC, Type~{\sc ii} (Type~B) represents the largest number among all GMC types, while it is Type~C in M33.
The percentage of Type~C GMCs in M33 (40--45\,\%) is also larger than in LMC (26\,\%).
Note that the clusters with ages of 10--30\,Myr in LMC are likely far away from their parent GMCs \citep{2001PASJ...53..985Y, 2009ApJS..184....1K} and thus the relative percentage is largely unknown.

In addition, these discrepancies can be partly explained by the different spatial resolution and observed line,
as the classification of  LMC GMCs have been done with 40\,pc resolution in CO($J=1-0$) \citep{2008ApJS..178...56F}. 
First, the GMCs in M33 are typically double the size of those in LMC. If a M33 GMC is composed of two different types of GMCs, for example Types~B and C, whose sizes are comparable to those in LMC, then the GMC would have been classified as Type~C. Therefore, the classifications will be affected.
Second, regarding the different tracers used,
the M33 GMCs identified from CO($J=3-2$) data represent the denser and/or warmer molecular gas which is more directly linked to massive SF,
while the LMC GMCs identified from CO($J=1-0$) data trace the bulk of the molecular gas.
In fact, in the case of M33 GMCs identified by using CO($J=1-0$) data, more than two-thirds have associated H{\sc ii} regions \citep{2003ApJS..149..343E}, which is a similar proportion in LMC \citep{2009ApJS..184....1K}.

\subsection{Physical Properties of GMC Types}
\label{sec3.4.1}
\begin{figure}
  \begin{center}
    \plotone{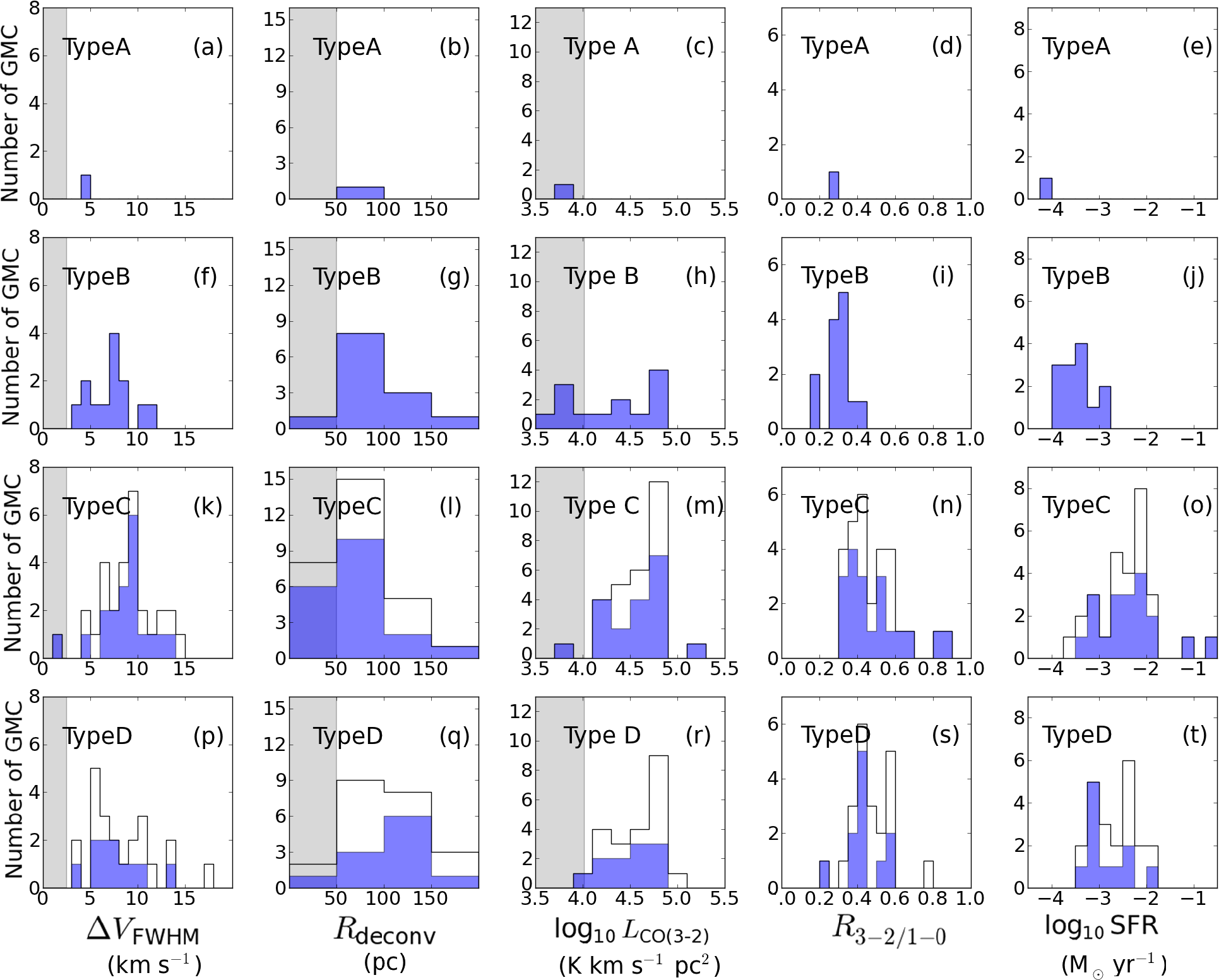}
   \end{center}
  \caption{Histograms of $\Delta V_{\rm FWHM}$, $R_{\rm deconv}$, $L^{\prime}_{\rm CO(3-2)}$, $R\ratio$ and SFR for the 65 GMCs. (a)--(d), (f)--(j), (k)--(o), and (p)--(t) panels show the properties of the GMC Type~A, Type~B, Type~C, and Type~D, respectively. The shaded areas indicate the detection limits: $\Delta V_{\rm FWHM}=2.5$\,km\,s$^{-1}$, $R_{\rm deconv}=50$\,pc, and $L_{\rm CO(3-2)}=10^4$\,K\,km\,s$^{-1}$\,pc$^2$. Note that $R_{\rm deconv}$ is not derived for 31 GMCs because the size cannot be derived for those with minor axis less than the beam size, and thus we set the non-deconvolved radius as the upper limit. The (blue) filled histogram indicates the selected 44 GMCs which are associated with the YSGs with confirmed ages (see details in Section~\ref{p3:sec534}). \label{p3:typeprop_fig}}
\end{figure}

Figure~\ref{p3:typeprop_fig} shows the distributions of the line width ($\Delta V_{\rm FWHM}$), deconvolved radius ($R_{\rm deconv}$), CO($J=3-2$) luminosity ($L^{\prime}_{\rm CO(3-2)}$), averaged line ratio ($R\ratio$) and SFR of the 65 GMCs. Each row represents each GMC type.
The shaded areas represent the histograms for the 44 selected GMCs with more accurate classifications (Section~\ref{p3:sec534}).
Table~\ref{p3:typeprop_tab} summarizes the mean and standard deviation of these variables for each GMC type.
The $R\ratio$ distributions of Type~C and D show a peak over 0.4, while those of Type~B are always smaller than the average value of the former types (i.e., $R\ratio$$< 0.3$).
The SFR distributions of Type~C and D show a peak over $2\times10^{-2}\,M_{\odot}\,{\rm yr}^{-1}$, while those of the other three types are smaller.
We have used Kolmogorov-Smirnov (K-S) tests to check whether the distributions for the different types differ. We cannot discard that the $\Delta V_{\rm FWHM}$, $R_{\rm deconv}$ and $L^{\prime}_{\rm CO(3-2)}$ distributions for these three types arise from the same distribution. However, the test indicates that $R\ratio$ and SFR distributions of Type C and D are different to those of Type B (p-value $<$ 0.01).
The selected GMCs with more accurate classification, as shown in the shaded histograms (Figure~\ref{p3:typeprop_fig}), indicate that the average values of all quantities for Type~C are the largest, followed by Type~D, Type~B, and Type~A, in this order.
This suggests that the Type~C GMCs are relatively large and show the most active SF among all types.
Note that all quantities of the single Type~A object are smaller than for the other three types. However, additional Type A sources need to be identified to properly characterize the properties of this class of objects.

In the following subsections, we present a close-up view of the CO($J=3-2$), CO($J=1-0$), $R\ratio$ and H$\alpha$ maps of the largest GMC in each GMC type, in order to illustrate the different properties among the four GMC types.
Figures~\ref{fig6-gmc-60} -- \ref{fig6-gmc-16} show for four GMCs representative of each type: (a) the CO($J=3-2$) integrated intensity map, (b) CO($J=1-0$) map, (c) the $R\ratio$ map, and (d) the H$\alpha$ images. GMC-60 is chosen as a representative of Type~A, GMC-8 of Type~B, GMC-1 of Type~C, and GMC-16 of Type~D.
The positions of H{\sc ii} regions, the 24\,$\micron$ sources, and the surface density map of young stars are also marked in these figures.

The CO($J=3-2$) emission peaks are in general slightly offset from the CO($J=1-0$) emission peak as we mentioned in Section~\ref{sec311}, but they are closer to the location of massive SF sites such as YSGs and H{\sc ii} regions.
These differences of spatial distributions necessarily results in a relatively high $R\ratio$ area in a GMC around such massive SF sites, and the gradient of $R\ratio$ that gradually increase in the direction toward such massive SF sites.

\subsubsection{GMCs without Massive SF Regions (Type~A)}
\label{p3:gmc60}
Figure~\ref{fig6-gmc-60} shows the images of GMC-60 of Type~A. No large variations in $R\ratio$ ($\sim 0.28$ on average over the GMC) or bright H$\alpha$ emission spots in this GMC are found. Besides it is a relatively small GMC compared to the average size and mass of all identified GMCs (Table~\ref{p3:typeprop_tab}).
Since only one GMC is classified as Type~A, it is unclear whether the Type~A source is representative of a class of similar sources or just a statistical outlier.
Additional Type~A sources need to be identified to properly characterize the properties of this class of objects.

\subsubsection{GMCs with H{\sc ii} Regions but not YSGs (Type~B)}
\label{p3:gmc8}
One-fifth of the GMCs in our catalog are classified as Type~B.
We cannot assure significant differences in $\Delta V_{\rm FWHM}$, size and $L^{\prime}_{\rm CO(3-2)}$ compared to those of Type~C and Type~D (see Figure~\ref{p3:typeprop_fig}).
However, the $R\ratio$ is smaller than the latter two types, with $R\ratio \sim 0.3$ on average, and never exceeding 0.5. In addition, the SFR is  also lower than the latter two types.
We find that all Type~B GMCs share a common trend, in the sense that higher $R\ratio$ inside GMCs are preferentially found close to H{\sc ii} regions and 24\,$\micron$ sources, although the peak of $R\ratio$ is not always coincident with the central position of H{\sc ii} regions.
No large spatial variation in $R\ratio$ over the GMC is found.

The associated H{\sc ii} regions and 24\,$\micron$ sources show a relatively random distribution over the GMC.
In addition, these sources are relatively small both in extent and in luminosity.
In fact, we find that they do not usually exceed $L({\rm H\alpha})\sim 10^{38}$\,erg\,s$^{-1}$ or $L({\rm 24\,\micron})\sim 10^{38.5}$\,erg\,s$^{-1}$, which corresponds to a single O7.5V star \citep{2005A&A...436.1049M}, assuming the conversion factor from ionizing photon rates and extinction-corrected H$\alpha$ luminosities in \citet{1996AJ....111.1252M}.
This suggests that the associated H{\sc ii} regions are ionized primarily by a star as massive as a O7.5V star, or later type of OB stars accompanied by several other less massive stars.

As an example of Type~B GMCs, we present a close-up view of GMC-8 in Figure~\ref{fig6-gmc-8}.
This GMC is known as the most massive GMC as traced by CO($J=1-0$) \citep{2003ApJS..149..343E,2007ApJ...661..830R}, but it is just identified in our CO($J=3-2$) catalog as the eighth brightest GMC.
The CO($J=3-2$) emission shows a more compact distribution than that of CO($J=1-0$) and the emission peaks coincide.
The $R\ratio$ slightly increases at the center of the GMC or close to H{\sc ii} regions and 24\,$\micron$ sources, but no large variation is found all over the GMC.

\subsubsection{GMCs with H{\sc ii} Regions and YSGs (Type~C)}
\label{p3:gmc1}
About half of the GMCs in our catalog are classified as Type~C.
The $\Delta V_{\rm FWHM}$, size and $L^{\prime}_{\rm CO(3-2)}$ of this type are similar to those in other types, but $R\ratio$ and SFR are the largest among all (Figure~\ref{p3:typeprop_fig} and Table~\ref{p3:typeprop_tab}).
There are gradients of $R\ratio$ around massive SF sites.
Peaks of $R\ratio$ do not coincide in general with the positions of these SF sites, as also found in Type~B GMCs.
The average $R\ratio$ of this type, $\sim$0.5, is usually larger than that of Type~B, and even exceeds over 1.0 at its peak.
In addition, unlike Type~B, we find that the associated H{\sc ii} regions usually exceed the equivalent $L({\rm H\alpha})$ of a single O7.5 star.

As an example of Type~C GMCs, we show the maps for GMC-1 in Figure~\ref{fig6-gmc-1}.
The CO($J=3-2$) emission shows a more compact distribution than that of CO($J=1-0$).
GMC-1 overlaps with two YSGs, YSG-73 (corresponding to the GHR NGC~604) and YSG-72.
The emission peak of CO($J=3-2$) is offset from that of CO($J=1-0$) and is located northward, closer to the largest YSG, YSG-73, and the center of the GHR.
A high $R\ratio$ ($>0.8$) area is found on the southern side of YSG-73, where compact H{\sc ii} regions are located \citep{1999ApJ...514..188C,2007ApJ...664L..27T}.
High $R\ratio$ ($>0.8$) preferentially around the GHRs suggests that these areas may be excited by young massive stars \citep{2007ApJ...664L..27T}. 
A gradient of $R\ratio$ extends from its peak to the south.
Panel (d) shows that the concentrations of many bright (massive) stars are located at the heart of the GHR.
Note that it is not certain whether the YSG-72 is physically associated with this GMC and NGC~604 because there is no spatial correlation between the YSG, H{\sc ii} region, and $R\ratio$.

\subsubsection{GMCs with H{\sc ii} Regions and 10--30\,Myr YSGs (Type~D)}
\label{p3:gmc16}
About one-third of the GMCs in our catalog are classified as Type~D.
Generally, the physical properties of this type is similar to Type~C (see Figure~\ref{p3:typeprop_fig} and Table~\ref{p3:typeprop_tab}).
We also find that the associated H{\sc ii} regions are slightly less bright than those in Type~C, but larger than those in Type~B.

Figure~\ref{fig6-gmc-16} shows the maps of GMC-16, a representative of Type~D GMCs.
The CO($J=3-2$) emission shows a similar distribution to CO($J=1-0$), and both CO peaks coincide.
Again, the peak of $R\ratio$ ($\sim0.4-0.6$) is preferentially enhanced in the vicinity of the associated massive SF sites but not always coincident with them.
Such relatively high $R\ratio$ area is elongated from north to south and there is a gradient of $R\ratio$ from the $R\ratio$ peak in the middle of the GMC to the both west and east edges of the GMC.

\section{Discussion}
\label{p3:disc}
\subsection{Evolution of the GMCs}
In this section, we discuss based on our results the possible evolution of a GMC by focusing on the dense gas formation occurring around the first generated stars, which presumably leads to SF.
We also estimate the typical lifetime of a GMC in M33.

Since there is no correlation between H{\sc ii} regions or YSGs,
we infer that Type~A is at an evolutionary stage before massive SF.
In the case of LMC with 40\,pc resolution and using the CO($J=1-0$) line \citep{2008ApJS..178...56F}, \citet{2009ApJS..184....1K} have shown that about a quarter of all GMCs are Type~A.  Possible reasons that so few GMCs in M33 are classified as Type~A are spatial resolution and sensitivity limit and the choice of the molecular gas tracer, as explained in Section~\ref{p3:sec534}.
Type~B GMCs, which are associated with H{\sc ii} regions but not with YSGs, are  interpreted as being at an evolutionary stage just after the formation of massive stars, such as a single O7.5V star or later OB type stars, that are still partly embedded to be seen in the optical as YSGs.
We suggest that Type~C GMCs being associated both with H{\sc ii} regions and very young stellar groups (less than 10\,Myr) are at an evolutionary stage with active massive star formation, as massive as earlier types than O7.5V stars.
Type~D GMCs associated with H{\sc ii} regions and relatively old (10--30\,Myr) stellar groups are considered to be at an evolutionary stage where they have been continuously forming massive stars for over at least 10\,Myr.

We find that $R\ratio$ is enhanced around the massive SF sites for Types~B, C and D GMCs (Section~\ref{p3:gmc8}--\ref{p3:gmc16}).
In order to roughly quantify the offset between them, we plot in Figure~\ref{p3:mindist} the number of pixels above a certain $R\ratio$ (0.1, 0.3, 0.5, 0.7 and 0.9) as a function of the distance to its closest YSG.
The average sizes of the 75 YSGs are shown in the same figure, for reference.
The dashed lines represent the frequency distribution expected if the same number of YSGs are distributed at random in the observed area.
It shows that the distance to the nearest YSG becomes smaller as $R\ratio$ increases, but it must be noted that the $R\ratio$ peaks and the nearest YSGs do not coincide in general. 

We can assume that $R\ratio$ is a good tracer of density when $R\ratio < 0.7$, if kinetic temperature is low ($\sim$20\,K), as explained in \citet[][]{2007PASJ...59...43M}. This suggests that dense molecular gas fraction is enhanced in the vicinity of previously formed YSGs.
For regions with high $R\ratio > 0.8$, the molecular gas can have a higher kinetic temperature due to the nearest GHRs and YSGs (within $\sim$50\,pc).
This is in good agreement with the suggestion by \citet{1997ApJ...483..210W} that the GHRs may raise the kinetic temperature of the neighboring ($<\,100\,$pc) molecular gas.
Therefore, the enhancement of dense molecular gas fraction around the previously formed massive stars imply that once SF starts, dense gas formation is favored around the SF sites, which will then lead to the next generation of SF.

Next we estimate a lower limit of the lifetime of a GMC from the age of the oldest stellar group contained within the GMC, assuming that the GMCs and YSGs are being formed in a nearly steady evolution.
If the classification is done based on the oldest stellar group within a GMC, the frequency distribution of each type is 1 (2\,\%), 13 (20\,\%), 19 (29\,\%), and 32 (49\,\%) out of 65 selected GMCs.
The YSGs associated with Type~C and Type~D GMCs are $8\,\pm\,4$\,Myr and $18\,\pm\,10$\,Myr old on average, respectively.
Therefore, we estimate that the timescale for each evolutionary stage is $8\,\pm\,4$ and $10\,\pm\,6$\,Myr in Type~C and Type~D, from the age difference between the associated YSGs.
If we further assume that the timescale for each evolutionary stage is proportional to the number of GMCs, then Types~A, B, C, and D are estimated to be $\sim$1, 3--7, 5--10, and 8--17\,Myr, respectively.
As a result, the typical lifetime (from Type~A to Type~D) of a GMC with masses of $\ga10^5\,M_{\odot}$ is roughly 20--40\,Myr.
Note that these results do not change significantly when we use only 44 GMCs with accurate ages for their YSGs (Section~\ref{p3:sec534}).

In LMC, the lifetime of a GMC with mass as small as $5\times10^4\,M_{\odot}$ was estimated to be 20--30\,Myr \citep{1999PASJ...51..745F,2001PASJ...53..985Y,2009ApJS..184....1K}.
The timescales of Types~A, B and C in LMC (corresponding to Types~{\sc i}, {\sc ii} and {\sc iii}) are 6\,Myr, 13\,Myr, and 7\,Myr, respectively. 
The timescales for Type~B in M33 are smaller than that in LMC, and for Type~C they are similar or slightly larger. It is not appropriate to compare the lifetime for Type~A in M33 and in LMC because of the limited number of GMCs in this type.
No classification exists for LMC GMCs after Type~C, thus we cannot compare (see Section~\ref{p3:sec534} for a discussion).
Taking into account that M33 GMCs are a few times larger than the LMC GMCs, M33 Type~C and D GMCs are possibly composed of smaller clouds. If that is the case, these small clouds in M33 Type~C and D may not be coeval.
This can explain the fact that the timescale of Type~C GMCs in M33 are slightly larger than in LMC.

As a conclusion, we propose a scenario as follows.
First, dense gas formation occurs at a certain place in a GMC.
After the molecular gas become dense enough to form stars (Type~A), massive SF occurs in such dense regions, forming H{\sc ii} regions (Type~B).
About 4--8\,Myr later, the first formed YSGs are not embedded anymore and 
become visible (Type~C).
Subsequent dense molecular gas formation occurs due to the effect of previously formed YSGs and the next generation stars are born in such a dense gas (Type~D).
In this way, the SF propagates from initially generated stars, until  the reservoir of molecular gas is exhausted.
Such a continuous SF can last at least 10--30\,Myr in a GMC with a typical mass of $\ga10^5\,M_{\odot}$ during their lifetime of 20--40\,Myr (after the dense gas formation).

\section{Summary}
\label{p3:sum}
We present a GMC catalog using  wide field (121\,arcmin$^2$ in total) and high sensitivity (1\,$\sigma=16$--32\,mK in $T_{\rm mb}$ for a velocity resolution of 2.5\,km\,s$^{-1}$) CO($J=3-2$) maps of the nearby spiral galaxy M33 taken with the ASTE 10-m dish, and a complementary new catalog of YSGs for which we have estimated the ages.

We summarize our main results as follows: 
\begin{enumerate}
\item We identify 71 CO($J=3-2$) GMCs from the CO($J=3-2$) data.  We discard from the analysis 6 GMCs which are at the edges of the observed field. The remaining 65 GMCs are characterized by the deconvolved sizes in the range of 12--157\,pc in radius and virial masses of $1.5\times10^4$--$8.9\times10^6$\,$M_{\odot}$, which are comparable to those of GMCs in our Galaxy.
\item We identify 75 YSGs from the excess of the surface density map of young stars that are associated with the identified GMCs. 
The total number of O stars in a YSG spans from 9--169, corresponding to total stellar masses of 10$^{3.5}$--10$^{4.7}\,M_{\odot}$, assuming that they have a Salpeter IMF.
The YSG ages are also estimated using the Padova model, and are found to be in the range of 4--31\,Myr (12\,Myr on average). These values are comparable to those of typical OB associations.

\item We compare the GMCs with the distribution of YSGs and classical H{\sc ii} regions from a compilation in the literature.
The identified 65 GMCs are successfully classified into four categories: Type~A showing no sign of massive SF, Type~B being associated only with relatively small H{\sc ii} regions, Type~C with both H{\sc ii} regions and young ($<$\,10\,Myr) stellar groups, and Type~D with both H{\sc ii} regions and relatively old (10--30\,Myr) stellar groups.
Out of the 65 GMCs, 1 (1\,\%), 13 (20\,\%), 29 (45\,\%), and 22 (34\,\%) are found to be Types~A, B, C, and D, respectively.

\item From a comparison of the distributions of the CO($J=3-2$)/CO($J=1-0$) intensity ratio, $R\ratio$, with the YSGs and H{\sc ii} regions, we find that the $R\ratio$ peaks within a GMC are slightly offset from the SF sites, but preferentially located around them.

\item We interpret the four types as representing an evolutionary sequence of the GMCs.
Assuming that the number of the GMC types is proportional to the timescale of each evolutionary stage, we roughly estimate that they are $\sim$1, 3--7, 5--10, and 8--17\,Myr, respectively for Types~A, B, C, and D. This yields that the lifetime of a GMC with mass of $\ga10^5\,M_{\odot}$ is 20--40\,Myr, similar or slightly larger than the lifetime of LMC clouds. 

\item Finally, we propose a scenario where after molecular gas becomes dense enough to form stars, massive SF occurs in such dense part. Then the molecular gas around the first generation stars becomes dense and forms the new generation of stars.
In this way, the SF propagates from the initially generated massive stars and continues during the GMC lifetime. 
\end{enumerate}

\acknowledgments
We thank the anonymous referee for a very useful report.
We gratefully acknowledge the contributions of the ASTE staff to the development and operation of the telescope.
The ASTE project is
driven by Nobeyama Radio Observatory (NRO) and ALMA-J/Chile Observatory, branches
of National Astronomical Observatory of Japan (NAOJ), in
collaboration with University of Chile, and Japanese institutes including University of Tokyo, Nagoya University, Osaka Prefecture University, Ibaraki University, Hokkaido University, Joetsu University of Education, Keio University, and Kyoto University. 
Observations with ASTE were in part carried out remotely from Japan by using NTT's GEMnet2 and its partner R\&E (Research and Education) networks, which are based on AccessNova in a collaboration between the University of Chile, NTT Laboratories, and NAOJ.

{\it Facilities:} \facility{ASTE, NRO:45m}.

\clearpage

\onecolumn

\begin{figure}
\epsscale{1.0}
  \begin{center}
         \plotone{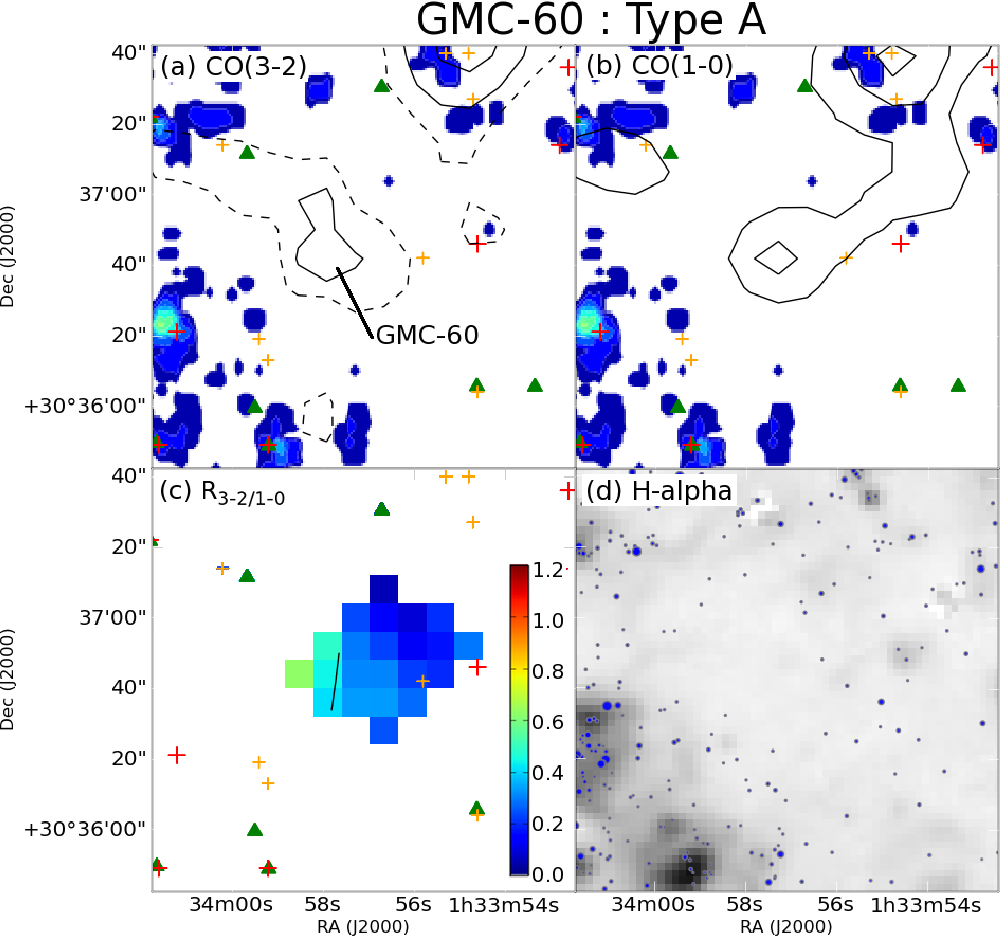}
 \end{center}
\caption{The distribution of {\bf (a)}  CO($J=3-2$) and {\bf (b)} CO($J=1-0$) emission for the single Type~A GMC identified in M33, GMC-60, overlaid on the surface density map of young stars (same as Figure~\ref{p3:fig3d}). The integrated velocity ranges from $V_{\rm LSR}=-171$ to $-161$\,km\,s$^{-1}$.
 Contour levels are 1, 2, 3, ..., 20\,K\,km\,s$^{-1}$. Dashed contours indicate 0.5 \,K\,km\,s$^{-1}$ level. Note that the contour levels below 2\,K\,km\,s$^{-1}$ are not shown for the CO($J=1-0$) map.
The red and orange cross signs, and green triangle signs are the same as in Figure~\ref{p3:fig3a}.
{\bf (c)} The $R\ratio$ map for GMC-60. Contour levels are 0.4, 0.6, 0.8, and 1.0. The color range spans from 0 to 1.2, and is shown on the right side of the panel.
{\bf (d)} The distribution of H$\alpha$ emission around GMC-60. 
The blue circles represent young stars which are selected using our criteria (Figure~\ref{p3:fig3e}a). Their sizes are proportional to the stellar brightness in $V$-band. 
\label{fig6-gmc-60}}
\end{figure}

\begin{figure}
  \begin{center}
  \epsscale{1.0}
         \plotone{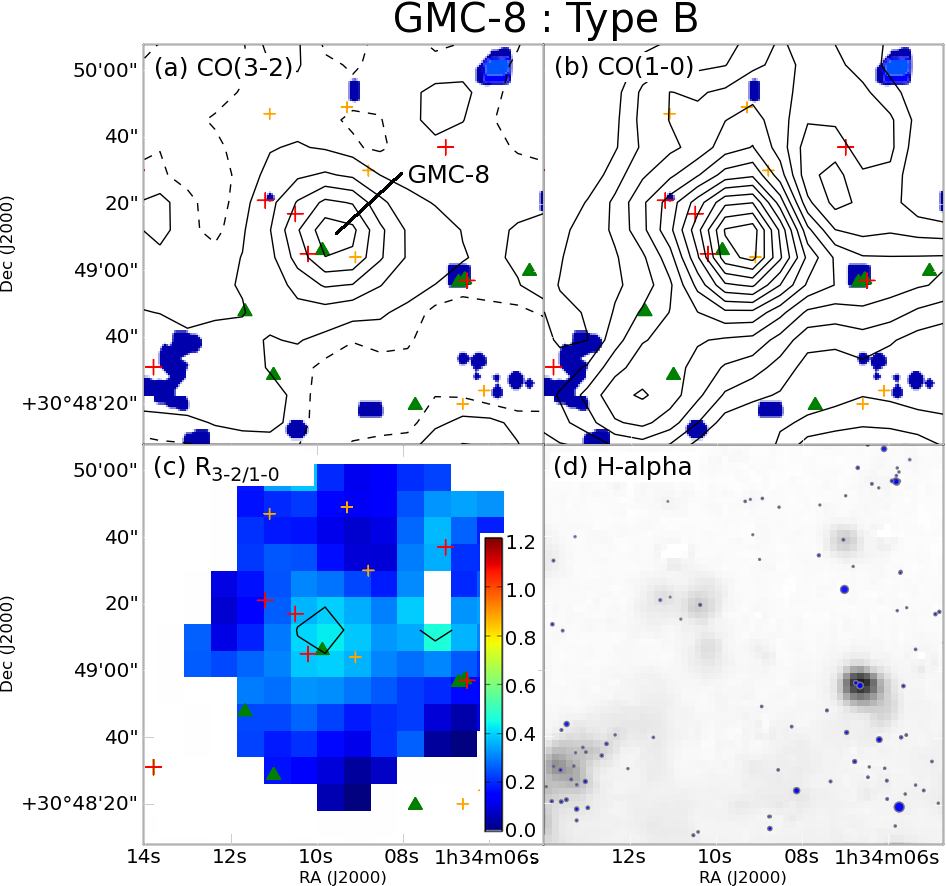}
 \end{center}
\caption{Same as Figure~\ref{fig6-gmc-60}, but for GMC-8 representing Type~B. The integrated velocity range used in the moment maps is $V_{\rm LSR}=-264$ to $-239$\,km\,s$^{-1}$.\label{fig6-gmc-8}}
\end{figure}

\begin{figure}
  \begin{center}
         \plotone{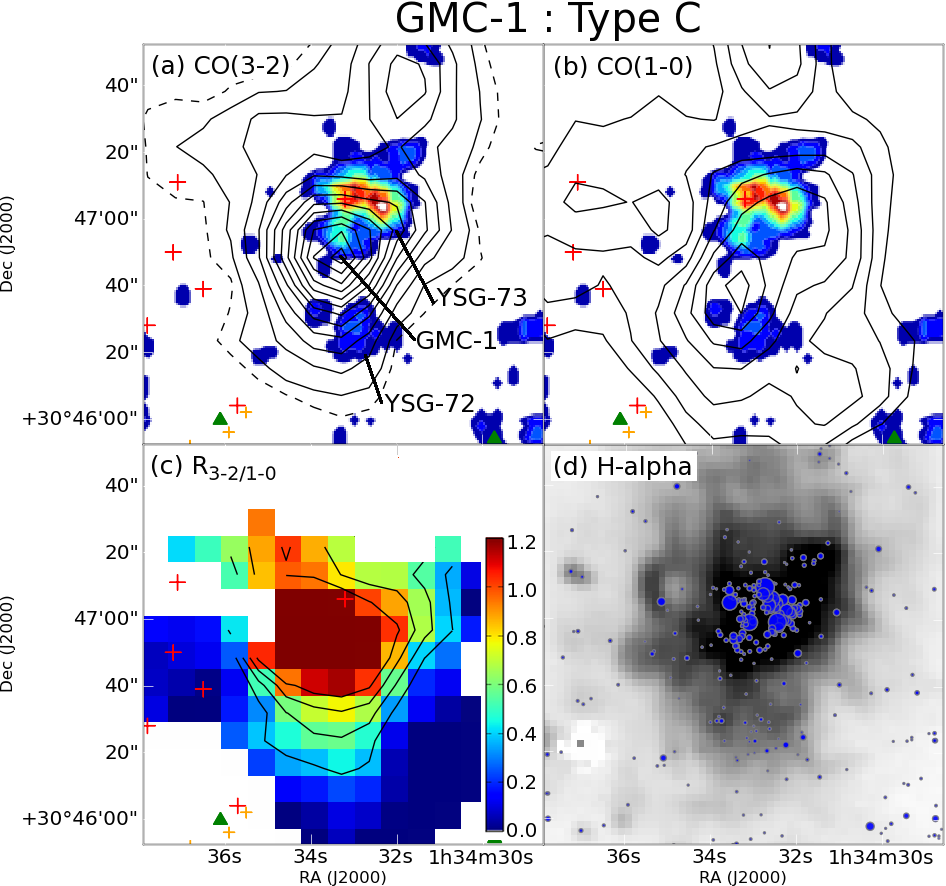}
 \end{center}
\caption{Same as Figure~\ref{fig6-gmc-60}, but for GMC-1 representing Type~C. The integrated velocity range used in the moment maps is $V_{\rm LSR}=-259$ to $-229$\,km\,s$^{-1}$.
 \label{fig6-gmc-1}}
\end{figure}

\begin{figure}
  \begin{center}
         \plotone{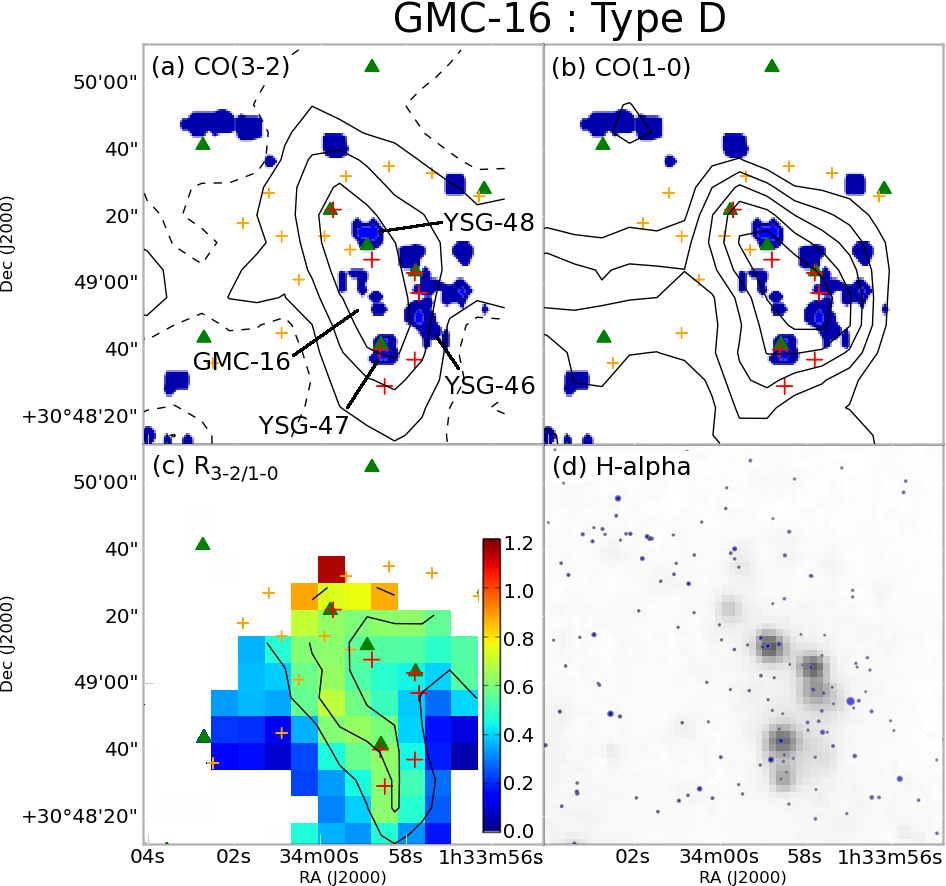}
 \end{center}
\caption{Same as Figure~\ref{fig6-gmc-60}, but for GMC-16 representing Type~D. The integrated velocity range used in the moment maps is $V_{\rm LSR}=-264$ to $-244$\,km\,s$^{-1}$. \label{fig6-gmc-16}}
\end{figure}

\begin{figure}
         \plotone{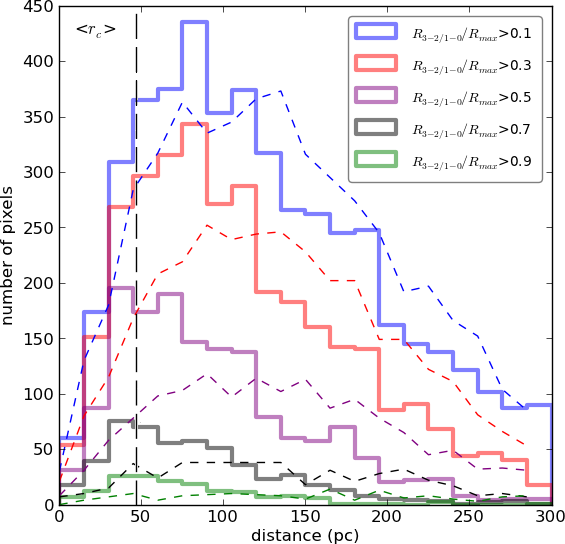}
\caption{Frequency distribution of the projected distances from the nearest YSG of the pixels where $R\ratio$ is above 0.1 (blue line), 0.3 (red), 0.5 (purple) , 0.7 (black) and 0.9 (green). 
For comparison, the dashed lines shows the frequency distribution of the distance when the YSGs are distributed randomly. 
The vertical (black) dashed line indicates the averaged size of the YSGs, for reference. \label{p3:mindist}}
\end{figure}

\clearpage

\begin{deluxetable}{ccccrc}
\tabletypesize{\scriptsize}
\tablecaption{Properties of the ASTE CO($J=3-2$) observations\label{p3:tab1}}
\tablewidth{0pt}
\tablehead{
\multicolumn{1}{c}{\textbf{Region No.}} & \colhead{\textbf{Map Center}} & \colhead{\textbf{Area}} & \colhead{\textbf{Int. Time}} & \colhead{\textbf{$\sigma_{\rm ch}$}}
\\
\colhead{} & \colhead{\textbf{R.A.,   Decl.}} & \colhead{\textbf{[arcsec, arcsec]}} & \colhead{\textbf{[hr]}} & \colhead{\bf [mK]} 
\\
\colhead{} & \colhead{\textbf{(J2000.0)}} & \colhead{} & \colhead{} & \colhead{} 
\\
\colhead{(1)} & \colhead{(2)} & \colhead{(3)} & \colhead{(4)} & \colhead{(5)} 
}
\startdata
1&{ 1$\h$33$\m$44\fs4}, { 30\arcdeg38\arcmin12\farcs2}& 310 $\times$ 335 & 22.5 & 16 
\\
2&{ 1$\h$34$\m$08\fs5}, { 30\arcdeg37\arcmin36\farcs4}& 263 $\times$ 240 & 8.7 & 19  
\\
3& { 1$\h$34$\m$12\fs8}, { 30\arcdeg34\arcmin28\farcs9}& 150 $\times$ 135 & 1.3 & 25  
\\
4& { 1$\h$33$\m$38\fs5}, { 30\arcdeg32\arcmin45\farcs0}& 200 $\times$ 440 & 21.8 & 18 
\\
5 & { 1$\h$33$\m$35\fs0}, { 30\arcdeg41\arcmin48\farcs0}& 120 $\times$ 120 & 1.9 & 32  
\\
6& { 1$\h$34$\m$05\fs5}, { 30\arcdeg48\arcmin02\farcs9}& 250 $\times$ 250 & 12.8 & 19 
\\
7&{ 1$\h$34$\m$33\fs2}, { 30\arcdeg47\arcmin06\farcs0}& 210 $\times$ 210 & 10.5 & 25  
\\
8& { 1$\h$34$\m$38\fs5}, { 30\arcdeg40\arcmin55\farcs0}& 180 $\times$ 150 & 7.0 & 18  
\\
\hline
\enddata
\tablecomments{
Col. (1): No. of the eight observed regions in order of increasing galaxy radius.
Col. (2): Central position of each region.
Col. (3): Size of the observed region.
Col. (4): Integration time for each region.
Col. (5): r.m.s. for a 2.5~\,km\,s$^{-1}$ channel, measured in an emission-free region.}
\end{deluxetable}

%
%
%

%
%
\begin{deluxetable}{cccccccccl}
\rotate
\tabletypesize{\scriptsize}
\tablecaption{Properties of the CO ($J=3-2$) emission components
\label{p3:co32cprops}}
\tablewidth{0pt}
\tablehead{
\multicolumn{1}{c}{\bf GMC ID}&
\multicolumn{1}{c}{\bf R.A., Decl.}&
\colhead{\bf $V_{\rm LSR}$}&
\colhead{\bf $T_{\rm mb}$}&
\colhead{\bf $\Delta V_{\rm FWHM}$}&
\multicolumn{1}{c}{\bf $A_{\rm maj}\times A_{\rm min}$}&
\colhead{\bf $L^{\prime}_{\rm CO(3-2)}$}&
\colhead{\bf $M_{\rm vir}$}&
\colhead{\bf $R\ratio$}&
\colhead{\bf Cross ID}\\
\multicolumn{1}{c}{}&
\multicolumn{1}{c}{\bf (J2000.0)}&
\colhead{(km\,s$^{-1}$)}&
\colhead{(K)}&
\colhead{(km\,s$^{-1}$)}&
\multicolumn{1}{c}{(pc$\times$pc)}&
\colhead{(10$^4$\,K\,km\,s$^{-1}$\,pc$^{2}$)}&
\colhead{(10$^5$\,M$_{\odot}$)}&
\colhead{}&

\\
\multicolumn{1}{c}{(1)}&
\multicolumn{1}{c}{(2)}&
\colhead{(3)}&
\colhead{(4)}&
\colhead{(5)}&
\multicolumn{1}{c}{(6)}&
\colhead{(7)}&
\colhead{(8)}&
\colhead{(9)} &
\colhead{(10)}
}
\startdata
1	&	1$\h$34$\m$33$\fs$2, 30$\arcdeg$46$\arcmin$52$\farcs$5	&	$-$243	&	0.84	&	13.8$\pm$2.8	&	72$\times$59	(42$^{\circ}$)			&		25.8$\pm$1.6		&		24.4$\pm$1.6		&	0.89$\pm$0.46	&		R07-124,	G11-215\\			
2	&	1$\h$33$\m$58$\fs$7, 30$\arcdeg$48$\arcmin$44$\farcs$5	&	$-$245	&	0.69	&	5.6$\pm$2.6	&	54$\times$46	(91$^{\circ}$)			&		8.3$\pm$6.1		&		\nodata		&	0.52$\pm$0.09	&		R07-86$^{\ast}$,	G11-242\\			
3	&	1$\h$34$\m$13$\fs$8, 30$\arcdeg$33$\arcmin$43$\farcs$9	&	$-$157	&	0.62	&	9.5$\pm$2.4	&	51$\times$49	(135$^{\circ}$)			&		8.5$\pm$1.4		&		2.0$\pm$0.3		&	0.52$\pm$0.12	&		R07-105\\						
4	&	1$\h$34$\m$10$\fs$8, 30$\arcdeg$36$\arcmin$15$\farcs$2	&	$-$160	&	0.60	&	6.6$\pm$1.1	&	65$\times$42	(83$^{\circ}$)			&		5.6$\pm$0.5		&		\nodata		&	0.45$\pm$0.11	&		R07-76\\						
5	&	1$\h$33$\m$52$\fs$5, 30$\arcdeg$39$\arcmin$19$\farcs$8	&	$-$168	&	0.59	&	10.0$\pm$1.7	&	59$\times$50	(29$^{\circ}$)			&		9.6$\pm$0.7		&		4.6$\pm$0.3		&	0.52$\pm$0.15	&		R07-1,	G11-108\\			
6	&	1$\h$33$\m$36$\fs$2, 30$\arcdeg$39$\arcmin$28$\farcs$5	&	$-$168	&	0.55	&	7.8$\pm$1.3	&	89$\times$47	(180$^{\circ}$)			&		9.4$\pm$1.0		&		\nodata		&	0.37$\pm$0.07	&		R07-34,	G11-105,	108\\		
7	&	1$\h$33$\m$33$\fs$7, 30$\arcdeg$41$\arcmin$25$\farcs$6	&	$-$185	&	0.53	&	9.9$\pm$2.4	&	60$\times$45	(150$^{\circ}$)			&		8.3$\pm$0.9		&		\nodata		&	0.84$\pm$0.36	&		R07-49\\						
8	&	1$\h$34$\m$09$\fs$4, 30$\arcdeg$49$\arcmin$07$\farcs$9	&	$-$249	&	0.50	&	10.4$\pm$1.8	&	55$\times$47	(137$^{\circ}$)			&		9.4$\pm$0.9		&		\nodata		&	0.26$\pm$0.03	&		R07-91,	G11-245\\			
9	&	1$\h$33$\m$35$\fs$5, 30$\arcdeg$36$\arcmin$29$\farcs$2	&	$-$135	&	0.48	&	8.5$\pm$4.3	&	63$\times$54	(5$^{\circ}$)			&		9.6$\pm$6.8		&		6.7$\pm$4.8		&	0.59$\pm$0.08	&		R07-37,	G11-87\\			
10	&	1$\h$33$\m$45$\fs$1, 30$\arcdeg$36$\arcmin$02$\farcs$1	&	$-$139	&	0.47	&	9.2$\pm$3.1	&	65$\times$54	(107$^{\circ}$)			&		9.4$\pm$4.0		&		7.9$\pm$3.3		&	0.56$\pm$0.11	&		R07-28,	G11-65\\			
11	&	1$\h$33$\m$49$\fs$6, 30$\arcdeg$33$\arcmin$55$\farcs$0	&	$-$134	&	0.46	&	5.5$\pm$1.1	&	125$\times$49	(161$^{\circ}$)			&		6.1$\pm$0.6		&		\nodata		&	0.40$\pm$0.06	&		R07-47,	G11-29,	30\\		
12	&	1$\h$34$\m$08$\fs$8, 30$\arcdeg$39$\arcmin$12$\farcs$4	&	$-$194	&	0.46	&	6.7$\pm$1.3	&	52$\times$49	(161$^{\circ}$)			&		5.7$\pm$0.6		&		1.1$\pm$0.1		&	0.34$\pm$0.07	&		R07-46,	G11-171\\			
13	&	1$\h$33$\m$29$\fs$9, 30$\arcdeg$31$\arcmin$59$\farcs$7	&	$-$133	&	0.46	&	9.2$\pm$2.1	&	62$\times$57	(174$^{\circ}$)			&		8.8$\pm$1.8		&		8.6$\pm$1.7		&	0.49$\pm$0.08	&		R07-82,	G11-35,	43\\		
14	&	1$\h$34$\m$06$\fs$8, 30$\arcdeg$47$\arcmin$43$\farcs$1	&	$-$255	&	0.45	&	8.6$\pm$1.1	&	88$\times$55	(87$^{\circ}$)			&		9.5$\pm$0.8		&		9.7$\pm$0.8		&	0.38$\pm$0.08	&		R07-81,	G11-256\\			
15	&	1$\h$33$\m$41$\fs$0, 30$\arcdeg$39$\arcmin$11$\farcs$5	&	$-$165	&	0.43	&	7.7$\pm$3.4	&	58$\times$47	(6$^{\circ}$)			&		5.7$\pm$3.6		&		\nodata		&	0.34$\pm$0.04	&		R07-22,	G11-103\\			
16	&	1$\h$33$\m$59$\fs$5, 30$\arcdeg$49$\arcmin$11$\farcs$6	&	$-$251	&	0.42	&	13.2$\pm$14.4	&	119$\times$62	(96$^{\circ}$)			&		12.1$\pm$27.5		&		35.2$\pm$80.0		&	0.55$\pm$0.09	&		R07-86$^{\ast}$,	G11-251\\			
17	&	1$\h$34$\m$02$\fs$5, 30$\arcdeg$39$\arcmin$08$\farcs$5	&	$-$201	&	0.41	&	9.2$\pm$1.9	&	63$\times$43	(81$^{\circ}$)			&		4.9$\pm$0.6		&		\nodata		&	0.41$\pm$0.07	&		R07-29,	G11-174\\			
18\tablenotemark{a}	&	1$\h$33$\m$59$\fs$4, 30$\arcdeg$36$\arcmin$01$\farcs$2	&	$-$152	&	0.40	&	7.0$\pm$2.2	&	63$\times$38	(109$^{\circ}$)			&		4.9$\pm$2.1		&		\nodata		&	0.72$\pm$0.20	&		R07-41\\						
19	&	1$\h$33$\m$34$\fs$7, 30$\arcdeg$37$\arcmin$00$\farcs$7	&	$-$136	&	0.39	&	5.4$\pm$6.9	&	76$\times$99	(99$^{\circ}$)			&		4.3$\pm$5.8		&		6.4$\pm$8.7		&	0.50$\pm$0.18	&		R07-38,	G11-88\\			
20	&	1$\h$34$\m$02$\fs$5, 30$\arcdeg$38$\arcmin$37$\farcs$6	&	$-$185	&	0.39	&	10.0$\pm$1.6	&	84$\times$46	(172$^{\circ}$)			&		8.3$\pm$0.7		&		\nodata		&	0.40$\pm$0.07	&		R07-31,	G11-170\\			
21	&	1$\h$33$\m$55$\fs$0, 30$\arcdeg$37$\arcmin$41$\farcs$9	&	$-$167	&	0.37	&	9.0$\pm$2.0	&	70$\times$66	(68$^{\circ}$)			&		7.0$\pm$1.4		&		11.8$\pm$2.3		&	0.39$\pm$0.07	&		R07-17,	G11-104\\			
22	&	1$\h$34$\m$15$\fs$5, 30$\arcdeg$39$\arcmin$11$\farcs$5	&	$-$191	&	0.37	&	8.9$\pm$0.9	&	134$\times$72	(168$^{\circ}$)			&		6.1$\pm$0.4		&		20.0$\pm$1.2		&	0.24$\pm$0.03	&		R07-66,77,	G11-182\\			
23	&	1$\h$33$\m$45$\fs$7, 30$\arcdeg$36$\arcmin$50$\farcs$6	&	$-$144	&	0.36	&	8.3$\pm$4.7	&	124$\times$72	(146$^{\circ}$)			&		10.8$\pm$7.7		&		16.6$\pm$11.9		&	0.44$\pm$0.07	&		R07-20,	G11-61,	62,	66	\\
24	&	1$\h$33$\m$47$\fs$3, 30$\arcdeg$32$\arcmin$52$\farcs$4	&	$-$118	&	0.35	&	12.5$\pm$2.0	&	105$\times$54	(114$^{\circ}$)			&		11.0$\pm$1.6		&		22.1$\pm$3.2		&	0.59$\pm$0.15	&		R07-61,	G11-23\\			
25	&	1$\h$34$\m$01$\fs$5, 30$\arcdeg$36$\arcmin$39$\farcs$6	&	$-$154	&	0.34	&	4.9$\pm$2.1	&	77$\times$44	(114$^{\circ}$)			&		3.3$\pm$1.8		&		\nodata		&	0.41$\pm$0.04	&		R07-42\\						
26	&	1$\h$34$\m$13$\fs$8, 30$\arcdeg$34$\arcmin$47$\farcs$6	&	$-$171	&	0.34	&	13.0$\pm$4.7	&	136$\times$50	(77$^{\circ}$)			&		12.5$\pm$3.8		&		17.5$\pm$5.4		&	0.59$\pm$0.13	&		R07-99\\						
27	&	1$\h$34$\m$34$\fs$5, 30$\arcdeg$46$\arcmin$21$\farcs$5	&	$-$222	&	0.33	&	5.8$\pm$2.9	&	56$\times$39	(92$^{\circ}$)			&		3.1$\pm$1.9		&		\nodata		&	0.55$\pm$0.17	&		R07-126,	G11-209\\			
28\tablenotemark{a}	&	1$\h$33$\m$56$\fs$7, 30$\arcdeg$40$\arcmin$10$\farcs$0	&	$-$193	&	0.32	&	8.9$\pm$1.3	&	67$\times$53	(47$^{\circ}$)			&		4.7$\pm$0.6		&		7.4$\pm$1.0		&	0.34$\pm$0.06	&		R07-7,	G11-175\\			
29	&	1$\h$33$\m$49$\fs$7, 30$\arcdeg$40$\arcmin$03$\farcs$3	&	$-$193	&	0.32	&	12.3$\pm$1.8	&	76$\times$66	(108$^{\circ}$)			&		6.3$\pm$0.8		&		23.9$\pm$3.0		&	0.38$\pm$0.07	&		R07-2,	G11-114\\			
30	&	1$\h$33$\m$33$\fs$4, 30$\arcdeg$32$\arcmin$16$\farcs$9	&	$-$130	&	0.32	&	14.3$\pm$4.6	&	89$\times$50	(34$^{\circ}$)			&		8.9$\pm$3.6		&		18.6$\pm$7.6		&	0.58$\pm$0.11	&		R07-73,	G11-42\\			
31	&	1$\h$33$\m$47$\fs$1, 30$\arcdeg$38$\arcmin$46$\farcs$4	&	$-$157	&	0.31	&	10.8$\pm$3.0	&	167$\times$63	(23$^{\circ}$)			&		11.7$\pm$3.9		&		28.8$\pm$9.7		&	0.59$\pm$0.08	&		R07-4,	G11-99,	100\\		
32	&	1$\h$33$\m$50$\fs$6, 30$\arcdeg$37$\arcmin$18$\farcs$2	&	$-$147	&	0.31	&	10.8$\pm$3.8	&	117$\times$77	(63$^{\circ}$)			&		9.8$\pm$3.9		&		28.7$\pm$11.4		&	0.55$\pm$0.10	&		R07-13,	G11-91,	92	\\	
33	&	1$\h$34$\m$32$\fs$0, 30$\arcdeg$47$\arcmin$38$\farcs$9	&	$-$248	&	0.30	&	10.8$\pm$5.1	&	50$\times$50	(77$^{\circ}$)			&		11.0$\pm$5.7		&		3.9$\pm$2.0		&	0.79$\pm$0.22	&		R07-123,	G11-217\\			
35	&	1$\h$34$\m$39$\fs$1, 30$\arcdeg$40$\arcmin$38$\farcs$9	&	$-$202	&	0.29	&	7.7$\pm$1.3	&	142$\times$79	(137$^{\circ}$)			&		9.3$\pm$1.2		&		17.0$\pm$2.1		&	0.28$\pm$0.04	&		R07-136$^{\ast}$\\						
36	&	1$\h$34$\m$03$\fs$2, 30$\arcdeg$46$\arcmin$34$\farcs$0	&	$-$245	&	0.29	&	11.8$\pm$2.4	&	127$\times$86	(136$^{\circ}$)			&		8.4$\pm$1.1		&		39.3$\pm$5.4		&	0.32$\pm$0.04	&		R07-62,	64,	G11-240,	244\\	
37	&	1$\h$33$\m$42$\fs$9, 30$\arcdeg$33$\arcmin$10$\farcs$9	&	$-$119	&	0.28	&	6.4$\pm$1.2	&	68$\times$55	(98$^{\circ}$)			&		4.8$\pm$0.6		&		4.4$\pm$0.6		&	0.40$\pm$0.08	&		R07-50,	G11-25\\			
38	&	1$\h$33$\m$40$\fs$7, 30$\arcdeg$37$\arcmin$34$\farcs$1	&	$-$145	&	0.28	&	17.3$\pm$3.1	&	189$\times$70	(49$^{\circ}$)			&		14.3$\pm$2.5		&		88.9$\pm$15.7		&	0.40$\pm$0.05	&		R07-16,	27,	G11-68,	69,	70\\
39	&	1$\h$33$\m$33$\fs$2, 30$\arcdeg$31$\arcmin$58$\farcs$9	&	$-$113	&	0.28	&	5.8$\pm$2.1	&	77$\times$46	(171$^{\circ}$)			&		3.6$\pm$1.4		&		\nodata		&	0.46$\pm$0.14	&		R07-75,	G11-36\\			
40	&	1$\h$33$\m$37$\fs$6, 30$\arcdeg$32$\arcmin$13$\farcs$8	&	$-$122	&	0.27	&	9.9$\pm$5.2	&	116$\times$58	(134$^{\circ}$)			&		8.2$\pm$5.6		&		17.5$\pm$12.1		&	0.38$\pm$0.05	&		R07-67,	71,	G11-37,	39\\	
41	&	1$\h$33$\m$52$\fs$9, 30$\arcdeg$38$\arcmin$59$\farcs$5	&	$-$150	&	0.27	&	7.2$\pm$1.6	&	59$\times$42	(149$^{\circ}$)			&		3.0$\pm$0.7		&		\nodata		&	0.51$\pm$0.12	&	G11-95\\						
42	&	1$\h$33$\m$48$\fs$1, 30$\arcdeg$39$\arcmin$31$\farcs$5	&	$-$168	&	0.25	&	9.2$\pm$4.8	&	70$\times$62	(157$^{\circ}$)			&		6.9$\pm$2.9		&		11.3$\pm$4.6		&	0.62$\pm$0.12	&		R07-3,	G11-107\\			
43	&	1$\h$33$\m$53$\fs$2, 30$\arcdeg$33$\arcmin$10$\farcs$4	&	$-$129	&	0.24	&	14.0$\pm$3.1	&	90$\times$74	(86$^{\circ}$)			&		9.4$\pm$1.6		&		39.2$\pm$6.8		&	0.39$\pm$0.07	&		R07-60\\					
44	&	1$\h$33$\m$52$\fs$7, 30$\arcdeg$37$\arcmin$11$\farcs$1	&	$-$151	&	0.24	&	6.7$\pm$5.2	&	94$\times$47	(94$^{\circ}$)			&		7.7$\pm$4.3		&		\nodata		&	0.41$\pm$0.05	&		R07-19$^{\ast}$,	G11-94\\			
45	&	1$\h$33$\m$52$\fs$8, 30$\arcdeg$37$\arcmin$41$\farcs$7	&	$-$157	&	0.24	&	6.0$\pm$4.6	&	78$\times$56	(84$^{\circ}$)			&		3.6$\pm$2.3		&		4.5$\pm$2.9		&	0.47$\pm$0.07	&		R07-19$^{\ast}$,	G11-98\\			
46	&	1$\h$33$\m$42$\fs$6, 30$\arcdeg$40$\arcmin$10$\farcs$9	&	$-$183	&	0.24	&	6.3$\pm$2.6	&	58$\times$42	(9$^{\circ}$)			&		2.3$\pm$0.8		&		\nodata		&	0.40$\pm$0.04	&	G11-112\\					
47\tablenotemark{a}	&	1$\h$34$\m$02$\fs$5, 30$\arcdeg$35$\arcmin$51$\farcs$8	&	$-$150	&	0.23	&	5.5$\pm$3.5	&	81$\times$56	(180$^{\circ}$)			&		3.3$\pm$3.2		&		3.9$\pm$3.8		&	0.61$\pm$0.12	&		R07-53\\				
48	&	1$\h$34$\m$36$\fs$4, 30$\arcdeg$40$\arcmin$20$\farcs$4	&	$-$192	&	0.22	&	7.5$\pm$1.7	&	105$\times$65	(68$^{\circ}$)			&		5.6$\pm$1.2		&		11.1$\pm$2.5		&	0.18$\pm$0.02	&		R07-136$^{\ast}$\\						
49	&	1$\h$33$\m$38$\fs$9, 30$\arcdeg$32$\arcmin$49$\farcs$7	&	$-$127	&	0.22	&	3.1$\pm$4.3	&	80$\times$63	(129$^{\circ}$)			&		2.3$\pm$3.2		&		1.5$\pm$2.1		&	$>$ 0.49	&	G11-41\\						
50	&	1$\h$33$\m$35$\fs$9, 30$\arcdeg$42$\arcmin$09$\farcs$7	&	$-$194	&	0.22	&	7.2$\pm$4.2	&	111$\times$56	(98$^{\circ}$)			&		5.0$\pm$3.1		&		8.4$\pm$5.3		&	0.42$\pm$0.11	&		\nodata\\				
51\tablenotemark{a}	&	1$\h$33$\m$33$\fs$4, 30$\arcdeg$37$\arcmin$06$\farcs$3	&	$-$142	&	0.22	&	5.3$\pm$3.5	&	97$\times$54	(84$^{\circ}$)			&		3.1$\pm$1.5		&		3.8$\pm$1.8		&	0.38$\pm$0.03	&		G11-89\\				
52	&	1$\h$34$\m$07$\fs$1, 30$\arcdeg$39$\arcmin$23$\farcs$7	&	$-$202	&	0.21	&	8.8$\pm$6.6	&	107$\times$62	(17$^{\circ}$)			&		2.4$\pm$1.3		&		14.6$\pm$8.1		&	0.39$\pm$0.06	&		R07-40,	G11-173\\			
53	&	1$\h$33$\m$23$\fs$5, 30$\arcdeg$31$\arcmin$58$\farcs$4	&	$-$121	&	0.21	&	11.3$\pm$1.8	&	80$\times$69	(110$^{\circ}$)			&		8.3$\pm$1.9		&		22.2$\pm$5.0		&	0.32$\pm$0.03	&		R07-90,	G11-40\\			
54	&	1$\h$34$\m$12$\fs$7, 30$\arcdeg$48$\arcmin$30$\farcs$3	&	$-$256	&	0.21	&	8.0$\pm$2.8	&	78$\times$62	(174$^{\circ}$)			&		9.1$\pm$2.8		&		9.5$\pm$2.9		&	0.33$\pm$0.05	&		R07-89,	93,	G11-254,	258	\\
55	&	1$\h$34$\m$04$\fs$7, 30$\arcdeg$48$\arcmin$59$\farcs$8	&	$-$251	&	0.20	&	4.9$\pm$2.2	&	126$\times$47	(152$^{\circ}$)			&		3.0$\pm$2.2		&		\nodata		&	0.32$\pm$0.02	&		R07-87,	G11-250\\			
56	&	1$\h$33$\m$43$\fs$8, 30$\arcdeg$38$\arcmin$57$\farcs$8	&	$-$150	&	0.20	&	6.0$\pm$3.2	&	82$\times$38	(60$^{\circ}$)			&		2.1$\pm$1.0		&		\nodata		&	0.33$\pm$0.03	&	G11-93\\						
57	&	1$\h$33$\m$57$\fs$9, 30$\arcdeg$46$\arcmin$35$\farcs$2	&	$-$246	&	0.20	&	4.6$\pm$1.7	&	57$\times$38	(47$^{\circ}$)			&		1.3$\pm$0.4		&		\nodata		&	$>$ 0.37	&	G11-235,	243\\			
58\tablenotemark{a}	&	1$\h$33$\m$51$\fs$4, 30$\arcdeg$35$\arcmin$53$\farcs$8	&	$-$146	&	0.19	&	16.1$\pm$6.3	&	47$\times$40	(84$^{\circ}$)			&		2.9$\pm$1.4	&			\nodata		&	0.48$\pm$0.09	&		G11-141\\				
59	&	1$\h$33$\m$45$\fs$5, 30$\arcdeg$33$\arcmin$20$\farcs$9	&	$-$129	&	0.19	&	6.4$\pm$3.1	&	93$\times$55	(136$^{\circ}$)			&		3.1$\pm$1.9		&		5.5$\pm$3.5		&	0.43$\pm$0.08	&	G11-28\\						
60	&	1$\h$33$\m$57$\fs$1, 30$\arcdeg$36$\arcmin$42$\farcs$4	&	$-$167	&	0.19	&	4.6$\pm$2.0	&	42$\times$42	(38$^{\circ}$)			&		1.2$\pm$0.7		&		\nodata		&	0.28$\pm$0.04	&		R07-30\\						
61	&	1$\h$33$\m$40$\fs$9, 30$\arcdeg$38$\arcmin$19$\farcs$1	&	$-$161	&	0.19	&	4.0$\pm$3.1	&	111$\times$102	(138$^{\circ}$)			&		2.4$\pm$2.0		&		4.8$\pm$4.1		&	0.38$\pm$0.38	&	G11-102\\						
62 &	1$\h$34$\m$13$\fs$3, 30$\arcdeg$47$\arcmin$14$\farcs$7	&	$-$245	&	0.18	&	8.2$\pm$2.4	&	69$\times$61	(109$^{\circ}$)			&		3.5$\pm$1.1		&		8.8$\pm$2.7		&	0.32$\pm$0.03	&	G11-207\\						
63	&	1$\h$34$\m$40$\fs$3, 30$\arcdeg$41$\arcmin$30$\farcs$6	&	$-$211	&	0.18	&	10.4$\pm$3.7	&	107$\times$80	(97$^{\circ}$)			&		5.9$\pm$1.4		&		26.3$\pm$6.4		&	0.34$\pm$0.03	&		R07-136$^{\ast}$\\						
64	&	1$\h$33$\m$50$\fs$7, 30$\arcdeg$33$\arcmin$47$\farcs$3	&	$-$126	&	0.17	&	3.1$\pm$2.2	&	97$\times$40	(169$^{\circ}$)			&		1.4$\pm$1.5		&		\nodata		&	0.43$\pm$0.10	&	R07-47,	G11-29	\\						
65	&	1$\h$34$\m$03$\fs$7, 30$\arcdeg$48$\arcmin$11$\farcs$3	&	$-$252	&	0.17	&	3.8$\pm$2.0	&	53$\times$53	(101$^{\circ}$)			&		1.2$\pm$1.1		&		0.9$\pm$0.9		&	0.32$\pm$0.03	&		R07-83,	G11-253\\			
66	&	1$\h$34$\m$10$\fs$8, 30$\arcdeg$37$\arcmin$08$\farcs$0	&	$-$166	&	0.16	&	6.5$\pm$1.9	&	50$\times$36	(59$^{\circ}$)			&		1.2$\pm$0.4		&		\nodata		&	0.18$\pm$0.02	&		R07-69,	G11-162\\			
67	&	1$\h$33$\m$44$\fs$8, 30$\arcdeg$32$\arcmin$06$\farcs$9	&	$-$106	&	0.16	&	11.9$\pm$6.4	&	65$\times$44	(156$^{\circ}$)			&		3.5$\pm$2.2		&		\nodata		&	$>$ 0.69	&			R07-68,	G11-21\\		
68	&	1$\h$34$\m$10$\fs$2, 30$\arcdeg$49$\arcmin$48$\farcs$5	&	$-$260	&	0.15	&	8.6$\pm$3.4	&	134$\times$52	(16$^{\circ}$)			&		3.6$\pm$1.4		&		10.1$\pm$3.9		&	0.26$\pm$0.04	&		R07-103,	G11-262\\			
69	&	1$\h$34$\m$02$\fs$5, 30$\arcdeg$48$\arcmin$39$\farcs$6	&	$-$255	&	0.15	&	5.3$\pm$4.0	&	46$\times$37	(64$^{\circ}$)			&		0.8$\pm$0.6		&		\nodata		&	0.29$\pm$0.04	&	G11-257\\						
70	&	1$\h$33$\m$59$\fs$6, 30$\arcdeg$37$\arcmin$12$\farcs$2	&	$-$171	&	0.14	&	7.9$\pm$3.6	&	50$\times$44	(71$^{\circ}$)			&		1.1$\pm$0.4		&		\nodata		&	0.43$\pm$0.16	&	G11-167\\				
71\tablenotemark{a}	&	1$\h$33$\m$22$\fs$6, 30$\arcdeg$32$\arcmin$14$\farcs$3	&	$-$141	&	0.13	&	16.6$\pm$5.4	&	47$\times$33	(93$^{\circ}$)			&		1.8$\pm$1.2		&		\nodata		&	\nodata	&	G11-136\\						
 \hline
\enddata
\tablecomments{
Col. (1): GMC ID in order of decreasing peak intensity in spectrum. 
Col. (2): Intensity-weighted peak position R.A., Decl..
Col. (3): Observed local standard of rest (LSR) velocity of the peak intensity.
Col. (4): Peak intensity in $T_{\rm mb}$
Col. (5): Full Width Half Maximum (FWHM) line width. 
Col. (6): Non-deconvolved major/minor axes and position angle in parenthesis (measured counterclockwise from west to north). 
Col. (7): CO($J=3-2$) luminosity. 
Col. (8): Virial mass. Note that virial masses of 27 GMCs are not shown when the minor axis of the GMC is smaller than the beam size.
Col. (9): Averaged $R\ratio$ over the extent of the GMC.
Col. (10): Cross identifications with other catalogs. R07-{\it n}: from \citet{2007ApJ...661..830R}. G11-{\it n}: from \citet{2012A&A...542A.108G}.
}
\tablenotetext{a}{GMCs at the edge of the field-of-view. The provided values are just lower limits.}
\tablenotetext{\ast}{A single GMC in the \citet{2007ApJ...661..830R}'s catalog appears to be composed of more than two GMCs in our catalog.}
\end{deluxetable}

%
%
%

%
%
\begin{deluxetable}{cccccccp{1.5in}}
\tabletypesize{\scriptsize}
 \tablecaption{M33 young stellar group catalog
 \label{p3:ys}}
\tablehead{
\textbf{YSG ID} &
\multicolumn{1}{c}{\bf R.A., Decl.}& 
\textbf{$r_{\rm cl}$} & \textbf{N$_{\rm O star}$}  &  \textbf{$M_v$} & \textbf{$E(B-V$)}& \textbf{Age}  & \textbf{Cross ID}\\ 
       & \multicolumn{1}{c}{\bf (J2000.0)}     &($\arcsec$) &       &  \textbf{(mag)} &&  \textbf{(Myr)}    &      \\ 
 (1) & (2) & (3) &(4)             & (5) & (6)               & (7)   & (8)
}
\startdata
1	&	$01\h33\m30\fs2$, $30\arcdeg31\arcmin44\farcs9$	&	8.3	&	21	&	17.62	&	0.10\tablenotemark{a}	&$	7	-	11	$&	SM464, PL12, SSA1272	\\
2	&	$01\h33\m33\fs6$, $30\arcdeg41\arcmin27\farcs6$	&	12.5	&	67	&	16.01	&	0.20\tablenotemark{a}	&$	3	-	8	$&	NGC595, PL57	\\
3	&	$01\h33\m34\fs3$, $30\arcdeg32\arcmin05\farcs8$	&	12.6	&	30	&	17.94	&	0.10	&$	6	-	12	$&	SSA1261, SSA1267, SSA1272	\\
4	&	$01\h33\m35\fs0$, $30\arcdeg37\arcmin02\farcs4$	&	7.4	&	17	&	18.84	&	0.10	&$	17	-	28	$&	SSA1284, SM113	\\
5	&	$01\h33\m35\fs2$, $30\arcdeg39\arcmin18\farcs1$	&	12.5	&	18	&	18.54	&	0.10	&$	11	-	17	$&	\nodata	\\
6	&	$01\h33\m35\fs7$, $30\arcdeg42\arcmin32\farcs2$	&	8.9	&	18	&	18.96	&	0.10	&$	14	-	25	$&	SSA1285	\\
7	&	$01\h33\m35\fs9$, $30\arcdeg36\arcmin29\farcs0$	&	9.8	&	24	&	18.38	&	0.10	&$	3	-	17	$&	SM128	\\
8	&	$01\h33\m36\fs4$, $30\arcdeg31\arcmin49\farcs1$	&	8.0	&	15	&	19.70	&	0.10	&$	22	-	39	$&	SSA1294	\\
9	&	$01\h33\m36\fs8$, $30\arcdeg36\arcmin35\farcs2$	&	8.0	&	14	&	18.05	&	0.10	&$	3	-	15	$&	\nodata	\\
10	&	$01\h33\m38\fs0$, $30\arcdeg32\arcmin01\farcs7$	& $<$	10.0	&	18	&	19.16	&	0.10	&$	3	-	35	$&	\nodata	\\
11	&	$01\h33\m38\fs7$, $30\arcdeg32\arcmin07\farcs9$	&	8.0	&	11	&	18.60	&	0.10	&$	6	-	19	$&	\nodata	\\
12	&	$01\h33\m39\fs3$, $30\arcdeg38\arcmin04\farcs0$	&	9.5	&	42	&	18.28	&	0.10	&$	3	-	17	$&	SSA1341, SSA1350	\\
13	&	$01\h33\m39\fs6$, $30\arcdeg32\arcmin37\farcs2$	&	17.2	&	67	&	18.11	&	0.10\tablenotemark{a}	&$	6	-	15	$&	SSA1343, SSA1345, PL139	\\
14	&	$01\h33\m40\fs5$, $30\arcdeg38\arcmin28\farcs0$	&	8.7	&	28	&	16.52	&	0.10	&$	5	-	8	$&	SSA1344 (SM545), SSA1361 (SM154)	\\
15	&	$01\h33\m42\fs2$, $30\arcdeg40\arcmin21\farcs5$	&	12.1	&	29	&	19.36	&	0.10	&$	22	-	28	$&	SSA1394, SSA1407	\\
16	&	$01\h33\m42\fs3$, $30\arcdeg33\arcmin00\farcs2$	&	9.3	&	22	&	17.64	&	0.10	&$	7	-	11	$&	SSA1406	\\
17	&	$01\h33\m43\fs1$, $30\arcdeg32\arcmin47\farcs6$	&	11.9	&	42	&	17.64	&	0.10	&$	7	-	11	$&	SSA1415	\\
18	&	$01\h33\m43\fs5$, $30\arcdeg39\arcmin05\farcs7$	&	17.0	&	65	&	16.40	&	0.10	&$	3	-	7	$&	SSA1430, SSA1416	\\
19	&	$01\h33\m44\fs3$, $30\arcdeg31\arcmin48\farcs0$	&	13.8	&	44	&	16.87	&	0.10	&$	3	-	8	$&	\nodata	\\
20	&	$01\h33\m44\fs8$, $30\arcdeg33\arcmin08\farcs2$	&	11.4	&	23	&	18.94	&	0.10	&$	19	-	35	$&	\nodata	\\
21	&	$01\h33\m44\fs9$, $30\arcdeg36\arcmin31\farcs1$	&	16.4	&	63	&	16.46	&	0.10	&$	3	-	7	$&	SM565	\\
22	&	$01\h33\m44\fs9$, $30\arcdeg35\arcmin53\farcs5$	&	9.9	&	25	&	16.78	&	0.10	&$	4	-	8	$&	\nodata	\\
23	&	$01\h33\m46\fs3$, $30\arcdeg36\arcmin52\farcs0$	&	14.4	&	42	&	15.64	&	0.10	&$	5	-	6	$&	SSA1472 (SM567)	\\
24	&	$01\h33\m46\fs7$, $30\arcdeg36\arcmin26\farcs9$	&	14.4	&	47	&	18.09	&	0.10	&$	10	-	17	$&	\nodata	\\
25	&	$01\h33\m47\fs5$, $30\arcdeg39\arcmin41\farcs8$	&	11.3	&	31	&	18.99	&	0.10	&$	14	-	22	$&	SSA1493	\\
26	&	$01\h33\m47\fs5$, $30\arcdeg38\arcmin36\farcs4$	&	9.0	&	30	&	18.23	&	0.10	&$	7	-	15	$&	SSA1495	\\
27	&	$01\h33\m48\fs1$, $30\arcdeg33\arcmin02\farcs3$	&	15.1	&	51	&	17.42	&	0.10	&$	6	-	12	$&	SSA1499	\\
28	&	$01\h33\m48\fs7$, $30\arcdeg38\arcmin45\farcs0$	&	8.0	&	17	&	19.25	&	0.10	&$	11	-	35	$&	SSA1515	\\
29	&	$01\h33\m48\fs7$, $30\arcdeg39\arcmin30\farcs0$	&	14.0	&	53	&	18.06	&	0.10	&$	5	-	17	$&	SSA1505 (SM572)	\\
30	&	$01\h33\m49\fs1$, $30\arcdeg39\arcmin45\farcs3$	&	13.4	&	59	&	16.30	&	0.10	&$	7	-	8	$&	SSA1516	\\
31	&	$01\h33\m49\fs3$, $30\arcdeg38\arcmin22\farcs5$	&	12.0	&	49	&	17.81	&	0.10	&$	8	-	14	$&	\nodata	\\
32	&	$01\h33\m50\fs1$, $30\arcdeg37\arcmin29\farcs6$	&	9.4	&	10	&	18.42	&	0.10	&$	3	-	22	$&	SSA1527	\\
33	&	$01\h33\m50\fs5$, $30\arcdeg33\arcmin46\farcs1$	&	10.0	&	12	&	18.95	&	0.10	&$	15	-	25	$&	SSA1536	\\
34	&	$01\h33\m51\fs2$, $30\arcdeg39\arcmin55\farcs8$	&	12.5	&	73	&	16.65	&	0.10	&$	5	-	7	$&	SSA1550	\\
35	&	$01\h33\m51\fs6$, $30\arcdeg37\arcmin22\farcs5$	&	10.0	&	20	&	19.79	&	0.10	&$	3	-	35	$&	\nodata	\\
36	&	$01\h33\m51\fs6$, $30\arcdeg38\arcmin51\farcs0$	&	12.2	&	84	&	17.15	&	0.10	&$	5	-	10	$&	SSA1554, SSA1546	\\
37	&	$01\h33\m52\fs3$, $30\arcdeg37\arcmin56\farcs0$	&	15.1	&	39	&	17.93	&	0.10	&$	8	-	12	$&	\nodata	\\
38	&	$01\h33\m52\fs6$, $30\arcdeg39\arcmin20\farcs3$	&	10.0	&	46	&	16.42	&	0.10	&$	3	-	7	$&	\nodata	\\
39	&	$01\h33\m52\fs8$, $30\arcdeg37\arcmin15\farcs0$	&	8.6	&	9	&	18.14	&	0.10	&$	3	-	15	$&	\nodata	\\
40	&	$01\h33\m53\fs0$, $30\arcdeg38\arcmin59\farcs4$	&	9.5	&	60	&	16.91	&	0.10	&$	3	-	8	$&	\nodata	\\
41	&	$01\h33\m54\fs5$, $30\arcdeg33\arcmin04\farcs3$	&	12.2	&	32	&	17.17	&	0.10\tablenotemark{a}	&$	7	-	11	$&	SSA1586 (SM215, PL167), SSA1598, SSA1585	\\
42	&	$01\h33\m55\fs6$, $30\arcdeg37\arcmin40\farcs0$	&	10.0	&	19	&	18.81	&	0.10	&$	12	-	25	$&	SM223	\\
43	&	$01\h33\m56\fs7$, $30\arcdeg40\arcmin20\farcs0$	&	10.5	&	15	&	19.36	&	0.01	&$	6	-	35	$&	SM236	\\
44	&	$01\h33\m57\fs4$, $30\arcdeg35\arcmin30\farcs0$	&	15.0	&	41	&	17.68	&	0.10	&$	12	-	14	$&	SSA1636 (SM241)	\\
45	&	$01\h33\m57\fs9$, $30\arcdeg48\arcmin46\farcs2$	&	8.9	&	11	&	18.98	&	0.15\tablenotemark{a}	&$	7	-	25	$&	PL81	\\
46	&	$01\h33\m58\fs6$, $30\arcdeg35\arcmin22\farcs5$	&	11.0	&	41	&	16.41	&	0.10	&$	3	-	7	$&	SSA1647 (SM476)	\\
47	&	$01\h33\m58\fs6$, $30\arcdeg48\arcmin37\farcs5$	&	10.0	&	13	&	18.74	&	0.10	&$	5	-	22	$&	SM253	\\
48	&	$01\h33\m59\fs0$, $30\arcdeg49\arcmin09\farcs2$	&	8.0	&	10	&	18.83	&	0.15\tablenotemark{a}	&$	14	-	28	$&	SM251 (PL80)	\\
49	&	$01\h33\m59\fs1$, $30\arcdeg35\arcmin37\farcs5$	&	10.0	&	32	&	17.51	&	0.10	&$	5	-	12	$&	\nodata	\\
50	&	$01\h33\m59\fs1$, $30\arcdeg35\arcmin47\farcs2$	&	10.0	&	28	&	19.13	&	0.10	&$	15	-	19	$&	SSA1654	\\
51	&	$01\h33\m59\fs2$, $30\arcdeg40\arcmin07\farcs5$	&	20.0	&	99	&	18.00	&	0.10	&$	6	-	15	$&	SSA1652	\\
52	&	$01\h34\m01\fs5$, $30\arcdeg36\arcmin20\farcs6$	&	10.8	&	38	&	17.61	&	0.10	&$	6	-	12	$&	SSA1693	\\
53	&	$01\h34\m01\fs6$, $30\arcdeg37\arcmin17\farcs0$	&	8.8	&	20	&	18.05	&	0.10	&$	3	-	15	$&	SSA1700, SSA1682	\\
54	&	$01\h34\m01\fs8$, $30\arcdeg46\arcmin28\farcs4$	&	16.4	&	31	&	18.10	&	0.10	&$	10	-	17	$&	SSA1694, SSA1685	\\
55	&	$01\h34\m02\fs2$, $30\arcdeg35\arcmin56\farcs0$	&	13.2	&	13	&	19.21	&	0.10	&$	3	-	15	$&	\nodata	\\
56	&	$01\h34\m02\fs3$, $30\arcdeg38\arcmin40\farcs6$	&	22.4	&	80	&	17.03	&	0.10	&$	5	-	7	$&	SSA1711 (SM276)	\\
57	&	$01\h34\m02\fs6$, $30\arcdeg37\arcmin04\farcs5$	&	8.3	&	9	&	18.38	&	0.10	&$	3	-	15	$&	\nodata	\\
58	&	$01\h34\m03\fs2$, $30\arcdeg46\arcmin36\farcs7$	&	9.2	&	9	&	18.50	&	0.10	&$	7	-	19	$&	SSA1715 (SM280)	\\
59	&	$01\h34\m03\fs3$, $30\arcdeg36\arcmin43\farcs6$	&	9.5	&	19	&	18.13	&	0.10	&$	3	-	15	$&	\nodata	\\
60	&	$01\h34\m05\fs7$, $30\arcdeg47\arcmin49\farcs8$	&	10.7	&	21	&	17.77	&	0.20\tablenotemark{a}	&$	3	-	14	$&	SM297 (PL197)	\\
61	&	$01\h34\m06\fs9$, $30\arcdeg47\arcmin22\farcs6$	&	13.2	&	30	&	16.44	&	0.20\tablenotemark{a}	&$	3	-	8	$&	SSA1748 (SM305, PL201)	\\
62	&	$01\h34\m07\fs6$, $30\arcdeg39\arcmin24\farcs4$	&	16.4	&	69	&	18.44	&	0.10	&$	14	-	25	$&	\nodata	\\
63	&	$01\h34\m09\fs3$, $30\arcdeg39\arcmin07\farcs7$	&	14.9	&	68	&	16.16	&	0.10\tablenotemark{a}	&$	3	-	6	$&	SSA1791, SSA1795, SSA1777 (SM321, PL98), PL97	\\
64	&	$01\h34\m11\fs5$, $30\arcdeg36\arcmin12\farcs2$	&	17.4	&	31	&	17.75	&	0.10	&$	8	-	14	$&	SSA1825	\\
65	&	$01\h34\m14\fs0$, $30\arcdeg33\arcmin39\farcs7$	&	7.2	&	16	&	17.95	&	0.10	&$	5	-	15	$&	SSA1862	\\
66	&	$01\h34\m14\fs7$, $30\arcdeg33\arcmin58\farcs5$	&	10.0	&	32	&	17.34	&	0.10	&$	8	-	12	$&	\nodata	\\
67	&	$01\h34\m15\fs1$, $30\arcdeg33\arcmin48\farcs1$	&	7.5	&	22	&	18.90	&	0.10\tablenotemark{a}	&$	3	-	25	$&	SM478 (PL27)	\\
68	&	$01\h34\m15\fs2$, $30\arcdeg35\arcmin00\farcs0$	&	8.6	&	14	&	17.89	&	0.10	&$	5	-	15	$&	SSA1894, SSA1895	\\
69	&	$01\h34\m16\fs0$, $30\arcdeg33\arcmin37\farcs5$	&	10.0	&	29	&	16.66	&	0.10	&$	3	-	7	$&	SSA1906 (SM480)	\\
70	&	$01\h34\m16\fs8$, $30\arcdeg39\arcmin18\farcs0$	&	8.3	&	10	&	19.76	&	0.10	&$	12	-	50	$&	\nodata	\\
71	&	$01\h34\m31\fs4$, $30\arcdeg47\arcmin48\farcs0$	&	9.5	&	11	&	19.12	&	0.10	&$	3	-	28	$&	\nodata	\\
72	&	$01\h34\m33\fs0$, $30\arcdeg46\arcmin23\farcs8$	&	10.4	&	21	&	19.07	&	0.10	&$	3	-	28	$&	\nodata	\\
73	&	$01\h34\m33\fs4$, $30\arcdeg47\arcmin03\farcs5$	&	21.2	&	169	&	15.90	&	0.10	&$	3	-	5	$&	NGC604	\\
74	&	$01\h34\m39\fs5$, $30\arcdeg41\arcmin37\farcs1$	&	8.7	&	16	&	18.73	&	0.10	&$	7	-	22	$&	SSA2167, SSA2168	\\
75	&	$01\h34\m40\fs1$, $30\arcdeg41\arcmin56\farcs0$	&	9.2	&	19	&	17.69	&	0.10	&$	3	-	14	$&	SSA2169, SSA2164	\\\hline  
\enddata
\tablecomments{
Col. (1): Young stellar group (YSG) ID, in order by right ascension . 
Col. (2): Central position R.A., Decl. of the YSG.
Col. (3): Size of the stellar group, which approximately corresponds to the boundary of $n^{\ast}=2$ stars per area (see Section~\ref{p3:5322}).
Col. (4): Number of O stars (more massive than O9V type star).
Col. (5): V magnitude of the brightest member in the YSG. 
Col. (6): Applied reddening correction for each YSG. 
Col. (7): Estimated age of the YSG. 
Col. (8): Cross identifications with other catalogs. SM-{\it n}: from \citet{2007AJ....134..447S}. PL-{\it n}: from \citet{2007AJ....134.2168P}. SSA-{\it n}: from \citet{2010ApJ...720.1674S}. NGC-{\it n}: giant H{\sc ii} regions.}
\tablenotetext{a}{From \citet{1999ApJ...517..668C}, \citet{2007AJ....134.2168P}, \citet{2000MNRAS.317...64G} and \citet{1999ApJ...514..188C}.}
\end{deluxetable}


\begin{deluxetable}{cccccccl}
\tabletypesize{\scriptsize}
 \tablecaption{Properties of M33 GMCs and massive star formation
 \label{p3:gmctype}}
\tablehead{
\textbf{GMC} & \textbf{Type} & \textbf{YSG}  & \textbf{$L({\rm H}\alpha)$}  & \textbf{$L(24\,\micron)$}  &\textbf{$A({\rm H}\alpha)$}  &\textbf{SFR} &
\multicolumn{1}{c}{\textbf{H{\sc ii} regions / $24\,\micron$ source}}\\ 
\textbf{ID} & \textbf{}& \textbf{ID} &(erg\,s$^{-1}$) &  (erg\,s$^{-1}$)  &(mag) & (M$_{\odot}$\,yr$^{-1}$)  &\\ 
 (1) & (2) & (3) &(4) & (5) & (6)  & (7)              &\multicolumn{1}{c}{ (8)}   
 }
\startdata
1	&	C\tablenotemark{b}	&	YSG-72, 73	&	7.67$\times10^{39}$	&	4.04$\times10^{40}$	& 0.20 &	2.16$\times10^{-1}$	&	NGC~604			\\
2	&	D	&	YSG-45, 47	&	4.16$\times10^{37}$	&	6.71$\times10^{38}$	&0.52&	3.58$\times10^{-3}$	&	B-005 (V-426), B-074, C-Z233A \\
3	&	C\tablenotemark{b}	&	YSG-65, 66, 67, 69	&	2.59$\times10^{38}$	&	1.96$\times10^{39}$	&0.27&	1.05$\times10^{-2}$	&	C-Z258 (V-115), B-066\\
4	&	D	&	YSG-64	&	3.80$\times10^{38}$	&	1.63$\times10^{39}$	&0.16&	8.73$\times10^{-3}$	&	B-709 (V-174), C-Z407 (V-195) \\
5	&	C\tablenotemark{b}	&	YSG-38, 40	&	3.48$\times10^{38}$	&	1.51$\times10^{39}$	&0.17&	8.09$\times10^{-3}$	&	C-Z160A (V-84)	\\
6	&	D\tablenotemark{b}	&	YSG-5	&		1.40$\times10^{37}$	&	2.13$\times10^{39}$	&2.08&	1.13$\times10^{-2}$	&	B-033, B-628, B-203 (V-280, V-273), H-634 (V-234) \\
7	&	C\tablenotemark{b}	&	YSG-2		&	2.89$\times10^{39}$	&	1.42$\times10^{40}$	&0.19&	7.60$\times10^{-2}$	&	NGC~595 (V-184)	\\
8	&	B\tablenotemark{b}	&	\nodata	&	2.17$\times10^{36}$	&	9.04$\times10^{37}$	&1.03&	4.81$\times10^{-4}$	&	C-Z322, C-Z324, B-647 (V-434), B-302 (V507), V-407	\\
9	&	C	&	YSG-7, 9&	8.54$\times10^{37}$	&	4.76$\times10^{38}$	&0.21&	2.55$\times10^{-3}$	&	C-Z112, C-Z214, V-168	\\
10	&	C\tablenotemark{b}	&	YSG-22&	6.94$\times10^{37}$	&	4.07$\times10^{38}$	&0.22&	2.18$\times10^{-3}$	&	C-Z219 (V-65), B-641, B-633 	\\
11	&	D\tablenotemark{b}	&	YSG-33&	1.16$\times10^{38}$	&	1.47$\times10^{38}$	&0.05&	8.01$\times10^{-4}$	&	B-047, B-014 (V-131)	\\
12	&	C\tablenotemark{b}	&	YSG-62, 63&	5.06$\times10^{37}$	&	3.94$\times10^{38}$	&0.28&	2.10$\times10^{-3}$	&	B-689 (V-266), B-702 (V-240), C-Z251 (V-271) \\
13	&	C	&	YSG-1	&	2.47$\times10^{38}$	&	2.13$\times10^{39}$	&0.31&	1.14$\times10^{-2}$	&	B-211, C-Z109, C-Z045 (V-70)	\\
14	&	C\tablenotemark{b}	&	YSG-60, 61	&	5.83$\times10^{38}$	&	2.01$\times10^{39}$	&0.13&	1.08$\times10^{-2}$	&	B-690, C-Z183, C-Z249			\\
15	&	C\tablenotemark{b}	&	YSG-18	&	1.10$\times10^{36}$	&	1.89$\times10^{38}$	&2.19&	1.01$\times10^{-3}$	&	C-Z010A, B-044, B-038 (V-210)\\
16	&	D\tablenotemark{b}	&	YSG-45, 48	&	4.04$\times10^{37}$	&	1.66$\times10^{38}$	&0.16&	8.90$\times10^{-4}$	&	B-703 (V-431), B-669 (V-427), B-696 (V-429), C-224 \\
17	&	C\tablenotemark{b}	&	YSG-56	&	2.49$\times10^{37}$	&	1.62$\times10^{38}$	&0.24&	8.66$\times10^{-4}$	&	C-Z405 (V-198)	\\
18	&	C\tablenotemark{a}	&	YSG-44, 46, 49, 50	&	2.19$\times10^{38}$	&	2.55$\times10^{39}$	&0.40&	1.36$\times10^{-2}$	&	C-Z001A (V-165),	V-199 \\
19	&	D\tablenotemark{b}	&	YSG-4	&	1.95$\times10^{38}$	&	1.05$\times10^{39}$	&0.20&	5.62$\times10^{-3}$	&	B-251 (V-193)	\\
20	&	C\tablenotemark{b}	&	YSG-56	&	3.80$\times10^{38}$	&	1.63$\times10^{39}$	&0.16&	8.73$\times10^{-3}$	&	C-Z405 (V-198)	\\
21	&	D\tablenotemark{b}	&	YSG-42	&	1.47$\times10^{37}$	&	1.89$\times10^{38}$	&0.43&	1.01$\times10^{-3}$	&	IC142			\\
22	&	D\tablenotemark{b}	&	YSG-70	&	1.61$\times10^{37}$	&	3.83$\times10^{38}$	&0.70&	2.04$\times10^{-3}$	&	C-Z274			\\
23	&	C\tablenotemark{b}	&	YSG-21, 23, 24	&	2.19$\times10^{38}$	&	1.14$\times10^{39}$	&0.20&	6.10$\times10^{-3}$	&	B-1002, B-060 (V-185),	V-223 \\
24	&	C	&	YSG-27	&	1.99$\times10^{38}$	&	1.07$\times10^{39}$	&0.20&	5.73$\times10^{-3}$	&	B-631 (V-127), B-019, V-113 \\
25	&	C	&	YSG-52, 53, 57, 59	&	7.26$\times10^{37}$	&	6.94$\times10^{37}$	&0.04&	3.81$\times10^{-4}$	&	B-672, C-Z245\\
26	&	D	&	YSG-68	&	9.78$\times10^{37}$	&	6.72$\times10^{38}$	&0.25&	3.59$\times10^{-3}$	&	C-Z322A (V-47), B-660, C-Z172 	\\
27	&	D	&	YSG-72	&	7.27$\times10^{38}$	&	1.64$\times10^{39}$	&0.09&	8.84$\times10^{-3}$	&	NGC~604			\\
28	&	D\tablenotemark{a}	&	YSG-43, 51&	3.10$\times10^{36}$	&	9.60$\times10^{37}$	&0.84&	5.11$\times10^{-4}$	&	H-744, H-731 H-740			\\
29	&	C\tablenotemark{b}	&	YSG-30, 34		&	1.23$\times10^{38}$	&	3.42$\times10^{38}$	&0.11&	1.84$\times10^{-3}$	&	C-Z150, B-064, C-Z144 (V-72)\\
30	&	C	&	YSG-3		&	1.46$\times10^{38}$	&	1.19$\times10^{39}$	&0.29&	6.35$\times10^{-3}$	&	C-Z210 (V-60) \\
31	&	D\tablenotemark{b}	&	YSG-26, 28, 31	&	2.35$\times10^{38}$	&	7.03$\times10^{38}$	&0.12&	3.78$\times10^{-3}$	&	B-017 (V-239)	\\
32	&	D	&	YSG-32, 35&	1.07$\times10^{38}$	&	8.91$\times10^{38}$	&0.30&	4.76$\times10^{-3}$	&	B-020 (V-201), B-637 \\
33	&	C\tablenotemark{b}	&	YSG-18	&	1.54$\times10^{38}$	&	7.35$\times10^{38}$	&0.18&	3.94$\times10^{-3}$	&	B-038 (V-210)	\\
34	&	D	&	YSG-71	&	1.04$\times10^{39}$	&	2.15$\times10^{39}$	&0.08&	1.16$\times10^{-2}$	&	NGC~604, B-652\\
35	&	B\tablenotemark{b}	&	\nodata		&	6.04$\times10^{37}$	&	4.35$\times10^{37}$	&0.03&	2.41$\times10^{-4}$	&	C-Z347, C-Z353			\\
36	&	D	&	YSG-54, 58&	6.17$\times10^{37}$	&	2.22$\times10^{38}$	&0.14&	1.19$\times10^{-3}$	&	C-Z291, B-081			\\
37	&	C	&	YSG-16, 17, 20		&	1.31$\times10^{38}$	&	5.76$\times10^{38}$	&0.17&	3.09$\times10^{-3}$	&	B-061, C-Z139, C-Z010			\\
38	&	D	&	YSG-12		&	2.04$\times10^{38}$	&	9.55$\times10^{38}$	&0.18&	5.11$\times10^{-3}$	&	B-612, B-058 (V-144), C-Z135, C-Z081 \\
39	&	C	&	YSG-3	&	2.77$\times10^{37}$	&	1.19$\times10^{39}$	&1.05&	6.34$\times10^{-3}$	&	C-Z210 (V-60), B-034	\\
40	&	D	&	YSG-8, 10, 11, 13	&	2.07$\times10^{37}$	&	2.56$\times10^{38}$	&0.42&	1.37$\times10^{-3}$	&	B-200, B-206	\\
41	&	C\tablenotemark{b}	&	YSG-36, 38, 40	&	2.18$\times10^{38}$	&	9.18$\times10^{37}$	&0.02&	5.24$\times10^{-4}$	&	C-Z160A (V-84), B-010	\\
42	&	C\tablenotemark{b}	&	YSG-25, 29, 30	&	9.66$\times10^{37}$	&	1.29$\times10^{39}$	&0.45&	6.88$\times10^{-3}$	& C-Z144 (V-72), B-037, C-Z221, B-066	\\
43	&	C	&	YSG-41	&	3.11$\times10^{38}$	&	1.24$\times10^{39}$	&0.15&	6.65$\times10^{-3}$	&	B-092 (V-97)	\\
44	&	C	&	YSG-39	&	6.61$\times10^{36}$	&	3.35$\times10^{37}$	&0.19&	1.79$\times10^{-4}$	&	C-Z227			\\
45	&	D	&	YSG-35, 37	&	1.01$\times10^{37}$	&	7.23$\times10^{37}$	&0.26&	3.86$\times10^{-4}$	&	B-1501			\\
46	&	D\tablenotemark{b}	&	YSG-15	&	3.28$\times10^{36}$	&	7.27$\times10^{37}$	&0.66&	3.87$\times10^{-4}$	&	H-601, H-598 (V-283)	\\
47	&	C\tablenotemark{a}	&	YSG-55	&	2.19$\times10^{38}$	&	7.03$\times10^{37}$	&0.01&	4.10$\times10^{-4}$	&	C-Z246A (V-122)	\\
48	&	B\tablenotemark{b}	&	\nodata	&	2.77$\times10^{36}$	&	2.59$\times10^{37}$	&0.33&	1.38$\times10^{-4}$	&	H-1145 (V-277), H-1146, H-1266 (V-281)	\\
49	&	D	&	YSG-13	&	1.21$\times10^{38}$	&	3.54$\times10^{38}$	&0.11&	1.90$\times10^{-3}$	&	B-050 (V-91)	\\
50	&	D\tablenotemark{b}	&	YSG-6		&	5.80$\times10^{37}$	&	1.36$\times10^{38}$	&0.09&	7.33$\times10^{-4}$	&	NGC~595			\\
51	&	B\tablenotemark{a}	&	\nodata	&	6.06$\times10^{37}$	&	7.85$\times10^{37}$	&0.05&	4.28$\times10^{-4}$	&	B-209, B-610,	V-182	\\
52	&	C\tablenotemark{b}	&	YSG-62, 63	&	5.73$\times10^{37}$	&	1.54$\times10^{38}$	&0.11&	8.29$\times10^{-4}$	&	C-Z182, B-689 (V-266)	\\
53	&	B\tablenotemark{b}	&	\nodata		&	3.89$\times10^{37}$	&	2.63$\times10^{38}$	&0.25&	1.41$\times10^{-3}$	&	C-Z041 (V-69)	\\
54	&	B\tablenotemark{b}	&	\nodata		&	4.12$\times10^{37}$	&	1.48$\times10^{38}$	&0.14&	7.94$\times10^{-4}$	&	B-665 (V-438), V-398, V-407	\\
55	&	B\tablenotemark{b}	&	\nodata	&	9.58$\times10^{37}$	&	6.81$\times10^{37}$	&0.03&	3.78$\times10^{-4}$	&	B-302 (V-507), V-420 \\
56	&	C\tablenotemark{b}	&	YSG-18	&	1.54$\times10^{38}$	&	7.35$\times10^{38}$	&0.18&	3.94$\times10^{-3}$	&	B-038 (V-210), B-051 \\
57	&	B\tablenotemark{b}	&	\nodata	&	8.55$\times10^{36}$	&	3.89$\times10^{37}$	&0.17&	2.08$\times10^{-4}$	&	C-Z212 (V-422), H-769 (V-385) \\
58	&	B\tablenotemark{a}	&	\nodata	&	6.26$\times10^{36}$	&	2.50$\times10^{37}$	&0.15&	1.34$\times10^{-4}$	&	C-Z146 (V-209)	\\
59	&	D\tablenotemark{b}	&	YSG-20	&	5.53$\times10^{37}$	&	1.26$\times10^{38}$	&0.09&	6.79$\times10^{-4}$	&	B-021, C-Z220, B-636			\\
60	&	A\tablenotemark{b}	&	\nodata	&	3.81$\times10^{35}$	&	1.67$\times10^{37}$	&1.06&	8.89$\times10^{-5}$	&				\\
61	&	C\tablenotemark{b}	&	YSG-12, 14	&	1.81$\times10^{38}$	&	9.55$\times10^{38}$	&0.20&	5.11$\times10^{-3}$	&	B-612, B-058 (V-144)\\
62	&	B\tablenotemark{b}	&	\nodata	&	1.09$\times10^{37}$	&	8.03$\times10^{37}$	&0.27&	4.29$\times10^{-4}$	&	C-Z299 (V-402)	\\
63	&	C	&	YSG-74, 75	&	2.68$\times10^{38}$	&	9.17$\times10^{38}$	&0.13&	4.92$\times10^{-3}$	&	B-0654, B0738			\\
64	&	D\tablenotemark{b}	&	YSG-33	&	1.02$\times10^{38}$	&	1.47$\times10^{38}$	&0.06&	7.99$\times10^{-4}$	&	B-014 (V-131)	\\
65	&	B\tablenotemark{b}	&	\nodata	&	1.94$\times10^{37}$	&	7.66$\times10^{37}$	&0.15&	4.11$\times10^{-4}$	&	C-Z242 (V-486)	\\
66	&	B\tablenotemark{b}	&	\nodata	&	5.25$\times10^{36}$	&	2.95$\times10^{37}$	&0.21&	1.58$\times10^{-4}$	&	H-931, H945, H-944, H-939			\\
67	&	C\tablenotemark{b}	&	YSG-19	&	9.20$\times10^{36}$	&	1.30$\times10^{38}$	&0.47&	6.93$\times10^{-4}$	&	B-614,	V-100	\\
68	&	B\tablenotemark{b}	&	\nodata	&	2.49$\times10^{35}$	&	4.65$\times10^{37}$	&2.27&	2.47$\times10^{-4}$	&	H-907, H-933			\\
69	&	B\tablenotemark{b}	&	\nodata	&	5.50$\times10^{35}$	&	2.82$\times10^{37}$	&1.17&	1.50$\times10^{-4}$	&	H-823,	V-502	\\
70	&	B\tablenotemark{b}	&	\nodata	&	3.08$\times10^{37}$	&	2.52$\times10^{38}$	&0.29&	1.35$\times10^{-3}$	&	H-793, V-227 \\
71	&	B\tablenotemark{a}	&	\nodata	&	1.59$\times10^{37}$	&	8.05$\times10^{37}$	&0.19&	4.31$\times10^{-4}$	&	C-Z040			\\
\hline
\enddata
\tablecomments{
Col. (1): GMC ID.
Col. (2): GMC classifications: Type A shows no signature of massive star formation; Type B is associated with H{\sc ii} region(s); Type C is associated with H{\sc ii} region(s) and less than 10\,Myr-old stellar group(s); and Type D is associated with H{\sc ii} region(s) and 10--30\,Myr-old stellar group(s).
Col.(3): Young stellar groups (YSGs) associated with the GMCs.
Col. (4): Luminosity of H$\alpha$, as measured over the extent of a GMC.
Col. (5): Luminosity of $24\,\micron$, as measured over the extent of a GMC.
Col. (6): The H$\alpha$ attenuation is calculated as $A_{\rm H\alpha}=2.5\, {\rm log} [1+0.038\,  L(24\, \micron)/L({\rm H\alpha})]$ \citep{2005ApJ...633..871C,2007ApJ...668..182P}. 
Col. (7): Star formation rate (SFR), derived from $L(24\,\micron)$ and $L({\rm H}\alpha)$ using the relation between the extinction-corrected H$\alpha$ line emission and the SFR in \citet{2007ApJ...666..870C}. 
Col. (8): Name of the associated H{\sc ii} regions : B-$XX$ from \citet{1974A&A....37...33B}; C-$XX$ from \citet{1987A&A...174...28C}; H-$XX$ from \citet{1999PASP..111..685H}; V-$XX$ from \citet{2007A&A...476.1161V}. 
The names within parenthesis indicate a counterpart of H{\sc ii} regions in the 24\,$\micron$ catalog. 
}
\tablenotetext{a}{GMCs at the edge of the field-of-views. The values just provide lower limits.}
\tablenotetext{b}{The selected GMCs with more accurate classification. See details in the text in Section~\ref{p3:sec534}. }
\end{deluxetable}


\begin{deluxetable}{ccrr}
\rotate
\tablecolumns{4}
\tablewidth{0pc}
\tablecaption{GMC classification and evolution \label{p3:class}}
\tablehead{
\colhead{GMC Type}    &  \multicolumn{1}{c}{Observed Signature} &   \colhead{Number of GMCs}  & \multicolumn{1}{c}{LMC\tablenotemark{a}} }
\startdata
A & No H{\sc ii} regions or young stellar groups & 1 (1\,\%)  & 46 (24\,\%)\\
B & With H{\sc ii} region(s), but no young stellar groups & 13 (20\,\%) & 96 (50\,\%)\\
C & With H{\sc ii} region(s) and young ($<$10-Myr) stellar group(s) & 29 (45\,\%) & 49 (26\,\%)\\
D & With H{\sc ii} region(s) and relatively old ($>$10-Myr) stellar group(s) & 22 (34\,\%) & \nodata \\
\enddata
\tablenotetext{a}{GMC Type in LMC \citep{2009ApJS..184....1K}. The definition of Types~A, B and C in our classification correspond to their Types~{\sc i}, {\sc ii} and {\sc iii}, respectively.}
\end{deluxetable}

\begin{deluxetable}{ccccccccccccccc}
\rotate
\tablecolumns{15}
\tablewidth{0pc}
\tablecaption{Physical properties of the GMC types \label{p3:typeprop_tab}}
\tablehead{
\colhead{GMC}    &  
\multicolumn{2}{c}{Line Width}  &\colhead{}&  
\multicolumn{2}{c}{Size}  &\colhead{}& 
\multicolumn{2}{c}{Luminosity} &\colhead{} &
\multicolumn{2}{c}{Ratio} &\colhead{}&
\multicolumn{2}{c}{SFR} \\
\cline{2-3} \cline{5-6} \cline{8-9} \cline{11-12} \cline{14-15}
\colhead{Type}    &  
\multicolumn{1}{c}{$\langle\Delta V\rangle$} & \multicolumn{1}{c}{$\sigma_{\Delta V}$} &\colhead{}&  
\multicolumn{1}{c}{$\langle R_{\rm deconv}\rangle$} & \multicolumn{1}{c}{$\sigma_{R_{\rm deconv}}$}  &\colhead{}& 
\multicolumn{1}{c}{$\langle L_{\rm CO(3-2)}\rangle$} & \multicolumn{1}{c}{$\sigma_{L_{\rm CO(3-2)}}$} &\colhead{} &  
\multicolumn{1}{c}{$\langle R\ratio\rangle$} & \multicolumn{1}{c}{$\sigma_{R\ratio}$} &\colhead{}&
\multicolumn{1}{c}{$\langle {\rm SFR} \rangle$} & \multicolumn{1}{c}{$\sigma_{\rm SFR}$} \\
\colhead{}    &  
\multicolumn{2}{c}{(km\,s$^{-1}$)} &\colhead{}&  
\multicolumn{2}{c}{(pc)} &\colhead{}& 
\multicolumn{2}{c}{(10$^4$\,K\,km\,s$^{-1}$\,pc$^2$)}&\colhead{} &  
\multicolumn{3}{c}{} &
\multicolumn{2}{c}{(10$^{-3}$$M_{\odot}\,{\rm yr}^{-1}$)} 
} 
\startdata
A\tablenotemark{a} & 4.6 & \nodata && $< 86$\tablenotemark{b} & \nodata && $0.8$ & \nodata &&0.28 &  \nodata && 0.09& \nodata\\
B & 7.2 & 2.2 &&  69\tablenotemark{b} & 36 &&  2.8 & 2.1 &&0.30 & 0.07 &&   0.51 & 0.43\\
C & 8.8 & 3.0 &&  64\tablenotemark{b} & 38 &&  4.4 & 2.8 &&0.49 & 0.14 && 14.26 & 40.42\\
D & 8.4 & 3.5 &&  83\tablenotemark{b} & 37&&  4.4 & 2.3 &&0.47 & 0.11 &&   3.60 & 3.49\\
\cutinhead{Selected\tablenotemark{c}}
A\tablenotemark{a} & 4.6 & \nodata && $< 86$\tablenotemark{b} & \nodata && $0.8$ & \nodata &&0.28 &  \nodata && 0.09& \nodata\\
B & 7.2 & 2.2 &&  69\tablenotemark{b} & 36 &&  2.8 & 2.1 &&0.30 & 0.07 &&   0.51 & 0.43\\
C & 8.6 & 2.8 &&  69\tablenotemark{b} & 36 &&  4.4 & 3.3 &&0.49 & 0.16 && 19.27 & 49.15\\
D & 7.6 & 2.7 &&  103\tablenotemark{b} & 32&&  3.9 & 2.1 &&0.43 & 0.09 &&   2.55 & 3.17\\
\enddata
\tablenotetext{a}{The values for the single GMC classified as Type~A (GMC-60) are presented for reference.}
\tablenotetext{b}{When the minor axis of a GMC is smaller than the beam size, the deconvolved radius cannot be derived, and then the non-deconvolved radius is used as upper limit.}
\tablenotetext{c}{When only GMCs with more accurate classifications are used (44 GMCs), 1 (2\,\%), 13 (30\,\%), 19 (43\,\%), and 11 (25\,\%) are classified as Types~A, B, C, and D, respectively. See details in Section~\ref{p3:sec534}.}
\end{deluxetable}

\clearpage
\appendix
\section{Spatial Comparison of GMCs, Stellar Groups and H{\sc ii} Regions for Individual GMCs}
Figures~18 -- 78 ({\it to be available online only}) show the spatial comparison of GMCs, $R\ratio$, YSGs and H{\sc ii} regions for the identified GMCs except those at the edge of the observed field, GMC-18, GMC-28, GMC-47, GMC-51, GMC-58 GMC-71, and also four GMCs that have been mentioned in Section \ref{sec3.4.1}. 

\clearpage

\end{document}